\documentclass[11pt]{article}

\usepackage{graphicx}
\usepackage{amssymb}
\usepackage{amsmath}
\usepackage{graphicx}
\usepackage{amstext}
\usepackage{esint}

\newcommand{\Omo}{\Omega_m^0}

\newcommand{\OLo}{\Omega_{\Lambda}^0}

\newcommand{\rmo}{\rho_{m}^0}

\newcommand{\rmr}{\rho_m}

\newcommand{\rR}{\rho_R}

\newcommand{\wm}{\omega_m}

\newcommand{\rL}{\rho_{\CC}}

\newcommand{\rLo}{\rho_{\CC}^0}

\newcommand{\CC}{\Lambda}

\newcommand{\f}{\tilde{f}}
\newcommand{\cM}{{\cal M}}
\newcommand{\cMd}{{\cal M}^2}

\newcommand{\bCC}{\bar{\CC}}
\newcommand{\bCDM}{\bar{\CC}{\rm CDM}}

\newcommand{\nueff}{\nu_{\rm eff}}

\newcommand{\xiR}{\xi'}
\newcommand{\rRo}{\rho_r^0}

\newcommand{\be}{\begin{equation}}
\newcommand{\ee}{\end{equation}}

\newcommand{\wCC}{\omega_\CC}

\newcommand{\Tgz}{T_{\gamma 0}}

\newcommand{\Oro}{\Omega_{r}^0}

\begin{document}

\hyphenation{cos-mo-lo-gists un-na-tu-ral-ly in-te-gra-ting
ne-gli-gi-ble e-xis-ten-ce con-vin-cing des-crip-tion ma-xi-mum}

\begin{center}

{\large \textsc{\bf THE $\bCDM$ COSMOLOGY:
FROM INFLATION TO DARK ENERGY THROUGH RUNNING $\CC$}} \vskip 2mm

 \vskip 10mm

\textbf{\large Joan Sol\`a}\,\,
\textbf{and}
\textbf{Adri\`{a} G\'omez-Valent}

\vskip 0.5cm

High Energy Physics Group, Dept. ECM Univ. de Barcelona,\\
Av. Diagonal 647, E-08028 Barcelona, Catalonia, Spain

\vskip0.15cm
and
\vskip0.15cm

Institut de Ci{\`e}ncies del Cosmos\\
Univ. de Barcelona, Av. Diagonal 647, E-08028 Barcelona

\vskip0.4cm

E-mails:    sola@ecm.ub.edu, adriagova@ecm.ub.edu
 \vskip2mm

\end{center}
\vskip 15mm

\begin{quotation}
\noindent {\large\it \underline{Abstract}}.
Perhaps the deepest mystery of our accelerating Universe in
expansion is the existence of a tiny and rigid cosmological
constant, $\CC$. Its size is many orders of magnitude below the
expected one in the standard model (SM) of particle physics. This is a
very welcome fact, namely if we care at all about our own existence
and fate. However, we do not have a minimally satisfactory
explanation for our good fortune and for the failing of the SM at
that crucial point. To start with, an expanding Universe is not
expected to have a static vacuum energy density. We should rather
observe a mildly dynamical behavior $\delta\Lambda(t)\sim R\sim
H^2(t)$ with the expansion rate $H$. At the same time, it is natural
to think that the huge value of the primeval vacuum energy (presumably connected to some grand unified theory) was
responsible for the initial inflationary phase. In the traditional
inflaton models such phase is inserted by hand in the early epoch of
the cosmic evolution, and it is assumed to match the concordance
$\CC$CDM regime during the radiation epoch. Here, instead, we
consider a class of dynamical vacuum models which incorporate into a
single  vacuum structure $\bCC(H)$ the rapid stage of inflation,
followed by the radiation and cold matter epochs, until achieving
our dark energy  Universe. The early behavior of such ``running
vacuum model'' ($\bar\CC$CDM) bares resemblance with Starobinsky's
inflation in the early Universe and is very close to the concordance
model for the entire post-inflationary history.  Most remarkably,
the inflationary period in the $\bar\CC$CDM terminates with ``graceful
exit'' and the large entropy problem can be solved. The model is
compatible with the latest cosmological data on Hubble expansion and
structure formation, and at the same time presents distinctive
observational features that can be tested in the near future.

\end{quotation}

\newpage

\tableofcontents

\newpage

\vskip 6mm

 \noindent \section{Introduction}
 \label{sect:Introduction}
When Einstein first introduced the cosmological term, $\CC$, in the
gravitational field equations, it was assumed to be constant and
positive\,\cite{Einstein1917}. Its value was naturally fixed to be
of the order of the critical density. The reasons why Einstein
introduced that term are well-known\,\footnote{See e.g.
Ref.\,\cite{JSP-CCReview2013} for a recent review.}, and at present
are of mere historical interest. However, there is one thing that
remains intact: its tiny value and the belief (at least within the
standard lore) that $\CC$ is a true constant of Nature. Certainly we
cannot disagree about its value since it has apparently been measured\,\cite{CosmoObservations1,CosmoObservations2}. It roughly remains of the same order of magnitude as in the time of Einstein, namely it
is of order of the critical density. Later on a nonvanishing $\CC$
was used to try to cure various astronomical problems, until it
became clear that its theoretical status was quite precarious since
it was realized that it was far too small by any standards in
particle physic units. This was the origin of the so-called
``cosmological constant (CC) problem''\,\cite{JSP-CCReview2013,CCproblem}. At this juncture people felt it was time to desperately find a theoretical
explanation for $\CC$ being very small, most likely zero. It was
thought that some symmetry would do nicely the job. Supersymmetry\,\cite{SUSY}
(SUSY), for instance, has been frequently cited
as of the time when  Wess and Zumino explored this theoretical possibility
-- more than forty years ago\,\cite{WessZumino74}.

But nowadays we know that the CC term, $\CC$, is nonzero and that SUSY (and for that matter any other extension of the standard model (SM) of particle physics)
is virtually impotent to explain the CC problem in any less
troublesome way than the SM itself. Both the SM and the MSSM\,\cite{MSSM} (the minimal supersymmetric version of the SM) have essentially the same
acute CC problem. This is of course very disturbing, specially for
the health of our cherished SM of particle physics, a model whose theoretical status became lately significantly augmented after the proclaimed discovery of the Higgs boson\,\cite{HiggsDiscovery}, the last missing piece to  ``crown'' the particle physics puzzle. The bare truth is that despite the tremendous effort and final triumph of the Higgs boson finding  \,\cite{SLWu2014}, the ``excessive success'' of the SM  makes
it  ``so simple yet so unnatural''\,\cite{Altarelli2014}. Without
extra physics the Higgs boson (and hence the entire backbone
structure of the SM) stays radiatively unprotected from the physics
of the high scales, namely those extremely large scales associated
to the grand unified theories (GUT's), which -- we should perhaps
recall at this point -- are not just there to provide aesthetical
ideas for the unification of the gauge couplings,  but are also badly
needed to explain the essential facts of inflation and baryogenesis,
for example.

Quite surprisingly, it is usually said that there is no trace of
physics beyond the SM. However this is not quite true: in fact, for all its particle physic success the SM is in
manifest disagreement with the most basic
cosmological observations. If we do not find at the
moment any obvious disagreement at the microphysical level (through
e.g.  $5\sigma$ deviations, or more, in some particle physics observable)
does not necessarily mean that the SM is a truly watertight theory,
since it makes a catastrophic prediction as to the value of the vacuum
energy and hence of the cosmological constant (see \cite{JSP-CCReview2013} for a more vivid account), let alone its complete inability to explain or to hint at the nature of the dark matter. The upshot is that some fundamental aspects of the SM of particle physics turn out to be in blunt disagreement with the standard (or ``concordance'') model of cosmology -- the so-called $\CC$CDM model. We could say that by finding the Higgs boson (or even ``a Higgs boson'', of whatever nature) we have certified experimentally the reality of the CC problem. The CC problem, therefore, is no longer a formal conundrum!

It is thus more than plausible to say that there is, or there must be,
physics beyond the SM, and also beyond the $\CC$CDM! What is not obvious at all, at least to us, is where it
lies the new physics that should rescue the two standard models, one of particle physics and the other of cosmology, from wreckage. We believe at least on one thing: wherever it hides and
whatever it be the new physics, it should provide a sober and sound
explanation for the nature of the vacuum energy and its connection
with the value of the cosmological constant. And we also believe
that this explanation should involve a dynamical cosmological term,
namely one capable of tracing the history of the Universe from its
origin till the present days in a single unified framework.

While the traditional ``explanation'' is the existence of a
nonvanishing and positive cosmological constant, $\CC$,  whose
energy density equivalent, $\rL=\CC/8\pi G$, is of order of the critical
density, this cannot be a truly convincing explanation, as in fact
an expanding Universe is not expected to have a static vacuum
\,\cite{JSP-CCReview2013}\,. A smoothly evolving vacuum energy
density $\rL(t)=\rL(H(t))$ that borrows its time-dependence from the
Hubble rate $H=H(t)$ --- taken as the natural dynamical variable in
the Friedmann-Lema\^{\i}tre-Robertson-Walker (FLRW) background ---
is not only a qualitatively more plausible and intuitive idea, but
is also suggested by fundamental physics, in particular by quantum
field theory (QFT) in curved space-time\,\cite{BooksQFTcurved}. This
more formal point of view, based on the renormalization group
approach, in which $\CC=\bCC(H)$ is a running quantity with the
cosmic expansion, has been introduced in
\cite{SS1999,SS2000,Cristina2003,Babic02,SS05,Fossil07,SS0809},  and recently
emphasized in\,\cite{JSP-CCReview2013,JSP2011,JSP2013}\,\footnote{Recent comprehensive analyses successfully testing  a wide class of dynamical vacuum models in the light of the recent observations are also available, see \cite{GoSolBas2014,SolGo2014}.}.
Furthermore, some recent applications of these ideas for a possible
description of the complete cosmic history have been put forward in
the literature\,\cite{H4model}. These vacuum models represent a
conceptually new approach that goes beyond the first
phenomenological approaches on time evolving cosmological constant,
cf.
\cite{OzerTaha87,FreeseET87,CarvalhoET92,Overduin88,Lima9694}. For alternative formulations and recent developments along these lines, see \cite{Bennie,Tommi2014,Maroto2014} and references therein.  In
general, the notion  of a variable vacuum energy has also been
entertained in the literature from different and interesting points
of view, also from the historical and philosophical
perspective\,\cite{Fahr2012,Fahr2011,Rugh2002}.

In the last years, and well within the tradition of the old Dirac's
ideas, an independent source of puzzling news has generated also a
lot of interest. Frequent hints that  the electromagnetic fine
structure constant $\alpha_{\rm em}$ and/or the proton mass might be
changing with the cosmic time (and locally in space) are reported in the
literature\,\cite{MurphyWebbFlaumbaum2003,Reinhold06} -- for
reviews, see e.g. \cite{FundamentalConstants,Calmet2014}. Theoretical models already exist in the literature trying to explain such phenomena, see e.g. \cite{BarrowMagueijo2014}. It is
tantalizing to conceive that, if $\alpha_{\rm em}$ can evolve with the
cosmic expansion, all of the fundamental ``constants'' should
change in time as well, including the gravity coupling and the masses of all the elementary particles. Recently, these
ideas have been linked to a possible time variation of the vacuum
energy as well\,\cite{FritzschSola2012,FritzschSola2014,Sola2014}, and this opens a line of thought well in the context of the dynamical vacuum energy in an expanding Universe\,\cite{JSP-CCReview2013}. Since the Newtonian coupling, $G_N$, determines the Planck mass $M_P=G_N^{-1/2}$, a possible time variation of it would be tantamount to say that $M_P$  slowly evolves with the cosmic expansion\,\cite{FritzschSola2014}.

In this presentation of the  ``running $\bCDM$ model'' (in fact a
class of models) we show that the idea of a dynamical vacuum in an expanding Universe can be the sought-for touchstone enabling a substantially improved  account of the entire cosmological history as compared to the concordance $\CC$CDM with rigid $\CC$ term. The content is as follows: In sections 2-3 we
summarize the theoretical framework for dynamical vacuum models. In
sections 4-5 we discuss inflation in the context of the $\bCDM$
model and address some important thermodynamical aspects of it, in
particular the large entropy problem. Subsequently, in Sect. 6, we compare
the early cosmological behavior of the  running $\bCDM$ model with
Starobinsky inflation in alternative formulations. The low-energy
implications of the $\bCDM$ class are explored in Sect. 7, where we
confront these models with the most recent cosmological data from
distant supernovae (SNIa), the cosmic microwave background (CMB)
anisotropies, the baryonic acoustic oscillations (BAOs) and the
input from structure formation. We show that the notion of a running
vacuum  can be, in principle, compatible with the current observations,
and suggest that traces of their dynamics should be testable in the
near future. In the final section we provide some discussion and our
conclusions.



\section{Dynamical vacuum in an expanding Universe}
\label{sect:DynamicalVacuum}

The idea that in an expanding Universe the cosmological term $\CC$
and Newton's constant $G_N$ could be variable with time is not new
and in fact it can be viewed as reasonable and even
natural. Dirac's ideas in the thirties\,\cite{Dirac1937,Dirac1938}
on the so-called ``large number hypothesis'' were seminal concerning
the possible time evolution of the gravitational ``constant'' $G_N$.
They were disputed by E. Teller\,\cite{Teller1948} and further
qualified by R.H. Dicke\,\cite{Dicke1957}. It also  triggered
subsequent speculations by G. Gamow\,\cite{Gamow1967} on the
possible variation of the fine structure constant. Since then the
subject has been in continuous evolution, and more and more
sophisticated experiments are being
designed to monitor the possible
time (and space) variation of the fundamental constants -- see e.g.
\,\cite{FundamentalConstants,Calmet2014} for reviews.

While in the old days the cosmological term, $\CC$, in Einstein's
equations may not have attracted a lot of attention regarding to its
potential time variability (setting aside occasional episodes
where some astrophysical observations had suggested this
possibility), it is natural to entertain this option in earnest when we cope with the full cosmological context.
This is especially so if we take into account that $\CC$ defines the
energy density parameter $\rL=\CC/(8\pi G_N)$, which is interpreted
as the vacuum energy density of the expanding
Universe\,\footnote{Ten years later after $\CC$ was introduced by
Einstein\,\cite{Einstein1917} to insure a static non-evolving
Universe, G. Lema\^\i tre\,\cite{Lemaitre1927} introduced a
nonvanishing $\CC$ to discuss his dynamical models of the expansion
of the Universe, strongly motivated by E. Hubble's observations
prior to their publication in 1929\,\cite{Hubble1929}. A few years later, in 1934, Lema\^\i tre
discussed for the first time\,\cite{Lemaitre1934} the interpretation
of the CC term as vacuum energy and its associated negative
pressure --  see e.g.\,\cite{Rugh2002,Luminet2011} for further historical
discussions.}. It should perhaps  be surprising if an accelerating Universe were to carry a static vacuum energy density throughout the
entire cosmic history\,\cite{JSP-CCReview2013}. A more natural possibility, which is perfectly compatible with the Cosmological Principle, is that $\CC=\CC(t)$ and hence $\rL=\rL(t)$.

From that point of view, it is instructive to consider the possible
modifications that may undergo the basic conservation laws if one
makes allowance for the time variability of the fundamental
gravitational parameters $G_N$ and $\CC$.  The Bianchi identity
satisfied by the Einstein tensor on the \textit{l.h.s.} of
Einstein's equations reads $\nabla^{\mu}G_{\mu\nu}=0$, where
$G_{\mu\nu}=R_{\mu\nu}-(1/2) g_{\mu\nu} R$. It follows that the
covariant derivative of the \textit{r.h.s.} of Einstein's equations
must be zero as well:
$\bigtriangledown^{\mu}\,\left(G_N\,\tilde{T}_{\mu\nu}\right)=0$,
where $\tilde{T}_{\mu\nu}\equiv T_{\mu\nu}+g_{\mu\nu}\,\rL $ is the
full energy-momentum tensor of the cosmic fluid composed of matter
and vacuum. Using the explicit form of the
Friedmann-Lema\^{\i}tre-Robertson-Walker (FLRW) metric, the
generalized conservation law emerging from this dynamical framework
reads
\begin{equation}\label{BianchiGeneral}
\frac{d}{dt}\,\left[G_N(\rmr+\rL)\right]+3\,G_N\,H\,(1+\wm)\rmr+3\,G_N\,H\,(1+\wCC)\rL=0\,,
\end{equation}
where $\wm=p_m/\rmr$ is the equation of state (EoS) for matter and
$\wCC=p_{\CC}/\rL$ is the EoS for the vacuum.  The Hubble rate $H$
dynamics ensues directly from Einstein's equations. We generalize
them for the case of dynamical vacuum (and restrict to the flat FLRW
metric):
 \begin{eqnarray}
 3H^2&=&8\pi G (\rmr+\rL)  \label{friedr1}\\
 2{\dot H}+3H^2&=&- 8\pi G (\wm\rmr+\wCC\rL)\;. \label{friedr2}
\end{eqnarray}
The overdot denotes derivative with respect to cosmic time $t$.  One
can easily check that only two of the equations
(\ref{BianchiGeneral})-(\ref{friedr2}) are independent. For example,
substituting (\ref{friedr1}) in (\ref{BianchiGeneral}) we arrive at
\begin{equation}\label{eq:dotH}
\dot{H}+4\pi\,G_N\,(1+\wm)\rmr+4\pi\,G_N\,(1+\wCC)\rL=0\,.
\end{equation}
One can check that this equation can also be obtained by combining
(\ref{friedr1}) and (\ref{friedr2}).

Let us note that the previous formulae remain valid
if we sum over all cosmic components (matter and vacuum).
However, in the particular but rather common situation where there
is a dominant matter component (e.g. cold matter or relativistic
matter), it is possible to obtain the evolution law for the Hubble
function solely in terms of the vacuum term and that dominant
material fluid. A simple calculation from the above equations leads
to
 \be
\label{eq:EHmonocomp}
\dot{H}+\frac32\,(1+\wm)\,H^2=4\pi\,G\left(\wm-\wCC\right)\,\rL\,.
\ee As advertized, in this equation we cannot sum over the matter
components, as $\wm$ in it stands for the EoS of the dominant
one. This equation will be useful in the next sections.

Up to this point we have assumed that the EoS of the vacuum component
is completely general. While the previous equations account for the general
situation (applicable even if $\wCC$ would correspond to a general
DE fluid), we shall henceforth adopt the simplest scenario
corresponding to the vacuum state, i.e. $\wCC=-1$, as being a
characteristic feature of the EoS of vacuum  even for time-evolving
$\rL=\rL(t)$.

In view of the above considerations, we next consider the following
cosmological scenarios beyond the concordance model:

{\bf Scenario I}: $\rL=\rL(t)$ is assumed variable, and $G_N=$const.
In this case, Eq.\,(\ref{BianchiGeneral}) implies
\begin{equation}
\dot{\rho}_{m}+3(1+\omega_{m})H\rho_{m}=-\dot{\rho}_{\Lambda}\,.
\label{frie33}
\end{equation}
Since we now have $\dot{\rho}_{\Lambda}\neq 0$ it means we permit
some energy exchange between matter and vacuum, e.g. through vacuum
decay into matter, or vice versa. Obviously if
$\dot{\rho}_{\Lambda}=0$ we recover
$\dot{\rho}_m+3\,H\,(1+\wm)\rmr=0$, i.e. the standard covariant
matter conservation law. Its solution in terms of the scale factor
is well-known:
\begin{equation}\label{solstandardconserv}
\rmr(a)=\rmo\,a^{-3(1+\wm)}\,.
\end{equation}

{\bf Scenario II}: $\rL=\rL(t)$ is again variable, but $G_N=G_N(t)$
is also variable. In contrast to the previous case, here we assume
matter conservation in the standard form (\ref{solstandardconserv}).
As a result the following conservation law ensues:
\begin{equation}\label{Bianchi1}
(\rmr+\rL)\dot{G}_N+G_N\dot{\rho}_{\CC}=0\,.
\end{equation}
In this setting the evolution of the vacuum energy density is
possible at the expense of a running gravitational coupling:
$\dot{G}\neq 0$.

{\bf Scenario III}: Here we keep $\rL=$const., but $G_N=G_N(t)$ is
again variable. Now we find:
\begin{equation}\label{dGneqo}
\dot{G}_N(\rmr+\rL)+G_N[\dot{\rho}_m+3H(1+\wm)\rmr]=0\,.
\end{equation}
In this case  matter is again non-conserved and the gravitational
coupling is running. Despite the vacuum energy is constant in this
scenario, such situation can mimic a form of dynamical dark energy
since it implies a different expansion rate\,\cite{BasSola13a}.

The above three generalized cosmological scenarios differ from the
concordance $\CC$CDM model, but can stay sufficiently close to it if
we consider the recent history of our Universe. Let us finally note
that the above dynamical vacuum models can be extended at high
energies for a successful explanation of the inflationary
Universe\,\cite{H4model}.

\section{$\bCC$CDM: a vacuum model for the complete cosmic history}\label{sect:RunningVacuum} Thus far we have
sketched some feasible frameworks for the time evolution of the the
vacuum energy and the gravitational coupling. Let us however note
that with the generalized conservation law (\ref{BianchiGeneral})
and Friedman's equation (\ref{friedr1}) is not possible to determine
the solution of the cosmological equations. We need some information
on the evolution laws $\rL=\rL(t)$ or $G=G(t)$. Such information has
been provided in the past based on some phenomenological ansatz (see
e.g. \cite{CarvalhoET92,OzerTaha87,FreeseET87,Overduin88} and
references therein). However, here we wish to focus on a class of
models motivated by the theoretical framework of QFT in curved
spacetime, see \cite{SS2000,Cristina2003,Babic02,SS05,Fossil07,SS0809,JSP2011,JSP2013} -- cf. also the review \cite{JSP-CCReview2013} and the long list of
references therein.

Recall that in  particle physics we have theories such as
QED or QCD  where the corresponding gauge coupling constants $g_i$
run with an energy scale $\mu$, i.e. $g_i=g_i(\mu)$. The scale $\mu$
is usually associated to the typical energy of the process.
Following the same line of thought we can think of $\rL$ as an
effective coupling sensitive to the quantum effects and thereby
running with an energy scale $\mu$ representative of the
cosmological evolution. In the context of QFT in a curved
background, where gravity is a classical external field, the quantum
effects that shift the value of $\rL$ are exclusively produced by
the loops of matter fields\,\cite{BooksQFTcurved}. Naturally $\mu$
should be fixed on physical grounds, but it is quite reasonable to
assume that the running of $\rL$  should be associated with the
change of the spacetime curvature because if there is no change in
the background geometry there is no cosmological evolution at all.
In the FLRW metric, this means that $\mu$ should be related with $H$ and
its time derivatives. We denote this association in the simplified
way $\mu\sim H$ (recall that $H$ has dimension of energy in natural units), but we will be more precise soon.

In view of the foregoing, the following form has been proposed for
the renormalization group (RG) equation for the vacuum energy
density of the expanding
Universe\,\cite{JSP-CCReview2013,SS2000}:
\begin{equation}\label{seriesLambda}
\frac{d\rL(\mu)}{d\ln\mu^2}=\frac{1}{(4\pi)^2}\left[\sum_{i}\,B_{i}M_{i}^{2}\,\mu^{2}
+\sum_{i}
\,C_{i}\,\mu^{4}+\sum_{i}\frac{\,D_{i}}{M_{i}^{2}}\,\mu^{6}\,\,+...\right]\,.
\end{equation}
In this expression, $M_{i}$ are the masses of the particles
contributing in the loops, and $B_{i},C_i,..$ are dimensionless
parameters. The RG equation (\ref{seriesLambda}) provides the rate
of change of the quantum effects on the CC as a function of the
scale $\mu$. Provided we are interested only on the dynamics of the
current Universe, we may cut off the series at the quadratic
contributions, i.e. only the ``soft-decoupling'' terms of the form
$\sim M_i^2\,\mu^2$ will be of significance. Notice that the $M_i^4$
terms are absent, as they would trigger a too fast running of the CC
term. As a matter of fact these effects are ruled out by the RG
formulation itself since only the fields satisfying $\mu>M_i$ are to
be included as active degrees of freedom. Being $\mu={\cal O}(H)$
(as indicated above) it is obvious that such condition cannot be currently satisfied by the SM particles -- see, however\,\cite{SS1999}. The leading effects on the
running of $\rL$ are, according to Eq.\,(\ref{seriesLambda}), of
order $M_i^2\mu^2\sim M_i^2\,H^2$, and hence it is dominated by the
heaviest fields. In the context of a typical GUT near the Planck
scale, these are the fields with masses $M_i\sim M_X\lesssim M_P$.
For instance, in Ref. \cite{Fossil07} a specific scenario is
described which is connected to the effective action of QFT in
curved spacetime.

Let us recall that because of the general covariance of the
effective action, among the list of possible terms emerging from the
quantum effects one expects only terms carrying an even number of
time derivatives of the scale factor. If expressed in terms of the
Hubble rate (which is the most convenient quantity to parameterize
the extra contributions), it amounts to terms of the form $H^2$,
$\dot{H}$, $H^4$, $\dot{H}^2$, $H^2 \dot{H}$ etc. In contrast, the
linear terms in $H$ (and in general any term with an odd number of
derivatives of the scale factor, such as $H^3$, $\dot{H}\,H$,
$\ddot{H}$ etc) are not expected since they would be incompatible
with the general covariance of the effective
action\,\cite{SS2000,Fossil07,SS0809,JSP2011,JSP2013}. Let us remark
that at low energies only the $H^2$ and $\dot{H}$ terms are relevant. The
higher order ones can however be important for the early
Universe\,\cite{H4model,Essay2014,Mavro2014,SUGRA2014,GoSol2014}.

As we have agreed, $\mu\sim H$ is the natural association of the RG-scale
in cosmology. However, a more general option is to associate $\mu^2$
to a linear combination of $H^2$ and $\dot{H}$ (both terms being
dimensionally homogeneous). Adopting this setting and integrating
(\ref{seriesLambda}) up to the terms of ${\cal O}(\mu^4)$  it is
easy to see that we can express the result as follows:
 \be
\rL(H,\dot{H})
=a_0+a_1\,\dot{H}+a_2\,H^2+a_3\,\dot{H}^2+a_4\,H^4+a_5\,\dot{H}\,H^2\,,
\label{GRVE} \ee
where the coefficients $a_i$ have different dimensionalities in
natural units. Specifically, $a_0$ has dimension $4$ since this is
the dimension of $\rL$; $a_1$ and $a_2$ have dimension $2$; and,
finally, $a_3$, $a_4$ and $a_5$ are dimensionless. The ``running
vacuum Universe'' ($\bCC$CDM) is the extension of the $\CC$CDM model
based on a dynamical vacuum energy density of the form (\ref{GRVE}), stemming from the basic RG equation (\ref{seriesLambda}).
While higher order term are still possible, that expression contains
the basic terms up to four derivatives of the scale factor, and
hence encodes the basic potential of the model both for the low and the high energy Universe.

Let us now consider a particularly simple and illustrative case of
the running $\bCC$CDM Universe. Suppose that rather than associating
$\mu^2$ with a linear combination of $H^2$ and $\dot{H}$ we would
just set $\mu^2=H^2$ (in this case the linear combination reduces to
just one term and we can just adopt the canonical choice $\mu=H$.).
In this situation we have $a_1=a_3=a_5=0$ in (\ref{GRVE}). The
remaining coefficients can be related immediately to those in
(\ref{seriesLambda}), and one can show that the final result can be
cast as follows:
\begin{equation}\label{lambdaH2H4}
\rL(H) = \frac{3}{8\pi G_N}\,\left(c_0 + \nu H^{2} +
\frac{H^{4}}{H_{I}^{2}}\right) \;,
\end{equation}
where $c_0$ has dimension $2$ and we have introduced the
dimensionless coefficient $\nu$ and the dimensionful one $H_I$.
Comparing with (\ref{seriesLambda}) it is easy to see that
\begin{equation}\label{eq:nualphaloopcoeff}
\nu=\frac{1}{6\pi}\, \sum_{i=f,b} B_i\frac{M_i^2}{M_P^2}\,.
\end{equation}
The dimensionful coefficient $H_I$ absorbs any other dimensionless
factor, but it is unnecessary to further specify its structure since
it represents a physical quantity (connected with the mechanism of
inflation) and can be determined (or at least bounded) by
observations, as we shall see in a moment. But let us first discuss
the interpretation of the dimensionless coefficient $\nu$. The sum
in (\ref{eq:nualphaloopcoeff}) involves both fermions and bosons.
Coefficient $\nu$ plays the role of the $\beta$-function coefficient
within the structure of the effective action in QFT in curved
spacetime. This is confirmed by the fact that $\nu$
depends on the ratio squared of the masses of the matter particles
to the Planck mass, which is indeed the expected result in
particular realizations of the RG in curved spacetime -- see e.g.
\cite{Fossil07}.  As we shall see, $H_I$ stands (to within very good
approximation) for the Hubble parameter in the inflationary epoch,
which is the only epoch where the higher order term $H^4$ can be of
relevance. During the inflationary period $H=$const. ($\dot{H}=0$),
so at least in the pure inflationary stage the terms we have dropped
from Eq.\,(\ref{GRVE}) should not be determinant. These terms,
however, can be important for the different modalities of reheating,
just after inflation.

Obviously the dynamical vacuum model (\ref{lambdaH2H4}) aims at
providing an unified description of the entire cosmic history, valid
from inflation to the present days. We will confirm if this is the
case in the subsequent sections. For the current Universe it is
enough to consider (\ref{lambdaH2H4}) up to the quadratic term, or type-A1 vacuum model:
\begin{equation}\label{lambdaTypeA1}
\rL(H) = \frac{3}{8\pi G_N}\,\left(c_0 + \nu H^{2}\right) \ \ \ \ \
\ \ \ \ \ ({\rm Type\ A1})\,.
\end{equation}
Note that if $\rLo$ and $H_0$ are the current values of the vacuum energy density and the Hubble parameter, then
\begin{equation}\label{lambdaTypeA1b}
c_0=\frac{8\pi\,G_N}{3}\,\rLo-\nu\,H_0^2\,.
\end{equation}
Taking into account that at low energy the terms $\sim\dot{H}$ can
be equally important as the $\sim H^2$ ones, one may also consider
the slightly extended model variant
\begin{equation}\label{lambdaTypeA2}
\rL(H) = \frac{3}{8\pi G_N}\,\left(c_0  +\tilde{\nu}\,\dot{H}+\nu
H^{2}\right) \ \ \ \ \ \ ({\rm Type\ A2})\,,
\end{equation}
where $\tilde{\nu}$ is a new dimensionless coefficient; it stems
from the original (dimensionful) $a_1$ in Eq.\,(\ref{seriesLambda}) after a
convenient redefinition. Although we use the same symbol $c_0$, it is understood that the connection of $c_0$ in (\ref{lambdaTypeA2}) with the current cosmological parameters is not exactly the same as in (\ref{lambdaTypeA1b}) but it can be obtained easily.

We will also introduce briefly two more vacuum
types (B1 and B2) for comparison. These are defined as follows:
\begin{eqnarray}
\rL(H) &=& \frac{3}{8\pi G_N}\,\left(c_0 + \epsilon H_0\,H\right)\ \ \ \ \ \ \ \ \  \ \ \ \ \ \ ({\rm Type\ B1})\label{lambdaTypeB1}\\
\rL(H) &=& \frac{3}{8\pi G_N}\,\left(c_0 + \epsilon H_0\,H+\nu
H^{2}\right)\label{lambdaTypeB2}\ \ \ \ \ ({\rm Type\ B2})\,,
\end{eqnarray}
where $\epsilon$ is a new dimensionless coefficient and $H_0$ is the
current value of the Hubble parameter. The dynamical vacuum models could
provide an alternative explanation for the dynamical DE in the
Universe  within the context of QFT in curved spacetime, thereby
representing an alternative to quintessence and other exotic DE
options. In the case of B1 and B2, however, the presence of the
linear term $\sim H$ (rather than $\sim\dot{H}$) is more
phenomenological because we do not expect linear terms in $H$ in the
effective action for the reasons explained above. Still these terms
can be admitted (they could represent bulk viscosity
effects\,\cite{BViscosity}, for instance) provided $c_0\neq 0$
and/or the standard terms (with an even number of derivatives of the
scale factor) are also present. In the case $c_0=0$ the above models
do not have a well-defined $\CC$CDM limit (i.e. none of them has in
this case a behavior near that of the concordance model), and one
can show that this is
problematic\,\cite{BasPolarSola12}. Recent analyses
have confronted these models to observations\,\cite{GoSolBas2014,SolGo2014} --
see also\,\cite{Cristina2003,SS05,BPS09,Grande2011,Gvariable,LXCDM} for
previous studies on a variety of possible scenarios. In Sect. \ref{sect:DynamicalVacuumCurrentUniverse} we provide a summarized presentation of
the main results for the most recent analyses.

\section{Inflation in the $\bCC$CDM model and Grand Unified Theories}\label{sect:InflationGracefulExit}

Let us now concentrate on the high energy implications of the
unified dynamical vacuum model.  To simplify our discussion we will
adopt the canonical form (\ref{lambdaH2H4}) since it already contains
the main features\,\footnote{The analysis of the modifications
introduced at high energies by the more general structure
(\ref{GRVE}) will be presented elsewhere.}. The Hubble function for
the dynamical vacuum models under consideration can be derived from
solving the basic equations of Sect. 2. However, these equations
depend on the particular scenario (I,II and III) we choose. For the present study we will focus on Scenario I of
Sect. 2 and hence on Eq.\,(\ref{frie33}).

\subsection{Solving the model in the early Universe}\label{sect:SolvingInflatonPhase}

Assuming that there is a dominant matter component in the early
Universe (typically radiation, $\wm=1/3$), we may combine equations
(\ref{eq:EHmonocomp}) and (\ref{lambdaH2H4}) and we arrive at the
following differential equation for $H$:
\begin{equation}
\label{HE} \dot
H+\frac{3}{2}(1+\omega)H^2\left[1-\nu-\frac{c_0}{H^2}-
\left(\frac{H}{H_I}\right)^{2}\right]=0.
\end{equation}
Now, if Eq.\, (\ref{lambdaTypeA1}) is to describe an approximate CC
term in the late Universe, the constant $c_0$ must be of the order
of the current value of the CC, or $\CC_0\simeq 3c_0$ to be more
precise, and at the same time the term $\sim\nu\,H^2$ can only
represent a small (dynamical) departure from it, thus $|\nu|\ll 1$.
This is not only what we expect, but in fact what we find when we
compare the low-energy models (\ref{lambdaTypeA1}) and
(\ref{lambdaTypeA2}) with the precision observational data collected
from the cosmological observations, we find indeed
$|\nu|,|\tilde{\nu}|\lesssim 10^{-3}$ (cf. Sect. 7).

In the early Universe, when $H$ is large, we can dismiss the term
$c_0/H^2\ll1$ in Eq.\,(\ref{HE}) to a very good approximation. This
implies the existence of a constant solution in that epoch, namely
\begin{equation}\label{eq:deSitterSolution}
H=\sqrt{1-\nu}\,H_I\simeq H_I\,,
\end{equation}
which corresponds to a de Sitter phase driven by a huge value of the
Hubble parameter $H_I$. We will corroborate that $H_I$ is the
characteristic value of the Hubble function of the high energy phase
of the early Universe, i.e. the effective value of $H$ in the
inflationary period, from our analysis of the energy densities
below. In what follows we will neglect $\nu$ in all practical
considerations concerning the early Universe. Notice that when
$H\simeq H_I$, the $\sim H^2$ term in Eq.(\ref{lambdaH2H4}) is
suppressed with respect to the $\sim H^4$ by precisely a factor
$|\nu|\ll1$. In the current Universe, where $H=H_0\sim 10^{-42}$ GeV
(the present value in natural units), the $\sim H^4$ term in
Eq.(\ref{lambdaH2H4}) is suppressed with respect to the $\sim H^2$
one by more than $100$ orders of magnitude! -- cf.
Eq.\,(\ref{eq:HIvalue}) below.

The value of the inflationary scale $H_I$ can be estimated from the
anisotropies of the cosmic microwave background as follows.
According to the CMB observations, the (primordial) spectrum ${\cal
P}_{\zeta}$ of the curvature perturbation\,\cite{LythLiddle} at
typical ``pivot'' length scales $k_0^{-1}$ (of a few tens to
hundreds of Mpc) is ${\cal P}_{\zeta}^{1/2}\simeq 5\times 10^{-5}$.
On the other hand, the theoretical calculation of ${\cal P}_{\zeta}$
is usually performed in the context of the inflaton model with
effective potential $V_{\rm eff}$, and yields
\begin{equation}\label{eq:Pzeta}
{\cal P}_{\zeta}(k_0)=\left.\frac{8}{3\,M_P^4}\,\frac{\langle V_{\rm
eff}\rangle}{\epsilon}\right|_{k_0}\,,
\end{equation}
where $\langle V_{\rm eff}\rangle$ is the vacuum expectation value
of the potential and $\epsilon$ is the standard slow roll
parameter\,\cite{LythLiddle}. Recall that $\epsilon$ is related to
the tensor-to-scalar fluctuation ratio $r\equiv n_T/n_s$ through
$r=16\epsilon$. Furthermore, the  value of the potential during
inflation is related to the GUT scale $M_X$ through $\langle V_{\rm
eff}\rangle\sim M_X^4$. For definiteness and simplicity we take them equal. Obviously this also defines the critical
density at the GUT epoch, $\rho_I=M_X^4$, which is related to $H_I$
by means of
\begin{equation}\label{eq:rhoI}
\rho_I=\frac{3H_I^2}{8\pi\,G_N}=M_X^4\,.
\end{equation}
From the previous equations we immediately find:
\begin{equation}\label{eq:MXvalue}
M_X=\langle V_{\rm
eff}\rangle^{1/4}=\left(\frac{3r}{128}\right)^{1/4}M_P\,{\cal
P}_{\zeta}^{1/4}=\left(\frac{r}{0.2}\right)^{1/4}\,2.2\times
10^{16}\,{\rm GeV}\,,
\end{equation}
and
\begin{equation}\label{eq:HIvalue}
H_I=\sqrt{\frac{8\pi}{3}}\,\frac{M_X^2}{M_P}\simeq
\left(\frac{r}{0.2}\right)^{1/2}\,1.1\times 10^{14}\,{\rm GeV}\,.
\end{equation}
Notice that we have normalized the tensor-to-scalar ratio to the
approximate value $r\simeq 0.2$ furnished by the BICEP2
collaboration\,\cite{BICEP2a} -- with the understanding that we are
waiting for a future confirmation/reanalysis of this important
experimental result. Notwithstanding, it is pretty clear that owing
to the fourth and squared roots involved of the factor $r/0.2$ in
the previous equations, the final estimates on $M_X$ and $H_I$ are
not critically sensitive to the precise value of $r$.

It is remarkable that the obtained value for $M_X$ in
(\ref{eq:MXvalue}) lies in the ballpark of $\sim 10^{16}$ GeV, which
is the expected order of magnitude in all viable GUT's compatible
with the current limits on proton decay. At the same time, the value
of $H_I$ obtained from (\ref{eq:HIvalue}) is seen to satisfy
\begin{equation}\label{eq:HIlimit}
\frac{H_I}{M_P}<10^{-5}\,,
\end{equation}
which is the condition that the fluctuations from the tensor modes
do not induce CMB temperature anisotropies larger than the observed
ones. Indeed, the vacuum fluctuations become classical a few Hubble
times after horizon exit, and have a spectrum $(H_I/2\pi)^2$
determined by the Gibbons-Hawing temperature $T_{GH}=H_I/2\pi$. To
be more precise, the primordial tensor perturbation $h_{ij}$ has the
spectrum
\begin{equation}\label{eq:TensorSpectrum}
{\cal
P}_h(k_0)=\left.\frac{64\pi}{M_P^2}\,\left(\frac{H_I}{2\pi}\right)^2\right|_{k_0}
=\left.\frac{16}{\pi}\,\left(\frac{H_I}{M_P}\right)^2\right|_{k_0}\,.
\end{equation}
Therefore, by requiring that ${\cal P}_{h}^{1/2}< 5\times 10^{-5}$ we
retrieve the bound (\ref{eq:HIlimit}) within the correct order of
magnitude from ${\cal P}_h$ alone.

Of course the previous considerations are just an estimate inasmuch
as they are partially based on the standard inflaton picture, which
is not exactly the one we are subscribing here. They nevertheless
lead to a reasonable estimate, for we could have just reversed the
sense of the argument and started from
Eq.\,(\ref{eq:TensorSpectrum}) rather than from (\ref{eq:Pzeta}).
Following this logic, we could have next applied the bound
(\ref{eq:HIlimit}) -- imposed by the measured anisotropies of the
CMB -- and derived an upper limit on the inflationary value of the
Hubble parameter directly from (\ref{eq:TensorSpectrum}), which is
under the constraint ${\cal P}_{h}^{1/2}< 5\times 10^{-5}$, and then
immediately derived the corresponding GUT scale (\ref{eq:MXvalue}).
The advantage of this way is that we know that the general form of
the tensor spectrum, $P_h\sim (H/M_P)^2$, is only sensitive to the
absolute value of the Hubble parameter squared -- measured in Planck
units -- at the inflation phase, a fairly general fact that holds
good irrespective of the underlying details of the inflation
dynamics. In this sense, the previous considerations are essentially
safe in so far as Eq.\,(\ref{eq:HIlimit}) is compatible
with them.

Far away from the inflationary period  (i.e. $H\ll H_I$) we find
another de Sitter solution of  Eq.\,(\ref{HE}), namely
$H=[c_0/(1-\nu)]^{1/2}$, whereby $\Lambda\approx 3\,c_0\sim
\Lambda_0$ (recall that $|\nu|\ll1$). This is of course the solution
we have mentioned before, which leads to the late time (approximate)
cosmological constant behavior, i.e. the current DE epoch. Let us
emphasize that the pure de Sitter phase of this late time solution
is achieved only in the remote future, but the fact that $\nu$ is
small but not exactly zero is responsible for a slow vacuum dynamics
$\sim H^2$ -- cf. e.g. Eq. (\ref{lambdaTypeA1}) -- in our recent
past (still going on at present) which could mimic a dynamical DE
scenario. This is precisely the situation when $\nu\neq0$
matters, and it can affect the present observations. We will address the details later (see Sect. 7), but first
we further analyze the inflationary phase.

\subsection{Achieving ``graceful exit'' in the $\bCC$CDM model}\label{sect:GracefulExit}

Let us now elucidate the cosmology of the $\bCC$CDM model in the
early de Sitter epoch.  We can virtually
set $c_0,\nu\to 0$ and assume $\omega=1/3$ since the subsequent matter epoch is presumably relativistic. It is convenient to trade the cosmic time for
the scale factor through $d/dt=aH d/da$. Eq.\,(\ref{HE}) takes
on the form
\begin{equation}\label{eq:Hea}
H'(a)+\frac{2}{a}\,H(a)\,\left(1-\frac{H^2}{H_I^2}\right)=0\,,
\end{equation}
where prime denotes differentiation with respect to the scale
factor. The solution of the previous equation renders
\begin{equation}\label{HS1}
 H(a)=\frac{H_I}{\left[1+D\,a^{4}\right]^{1/2}}\,.
\end{equation}
Substituting the above result in Eq. (\ref{lambdaH2H4}) we get the
explicit form of the vacuum energy density in terms of the scale
factor:
\begin{equation}\label{eq:rLa}
  \rho_\Lambda(a)={\rho}_I\,\frac{1}{\left[1+D\,a^{4}\right]^{2}}\,.
\end{equation}
The matter energy density then ensues after inserting the two
previous results in Friedmann's Eq.\,(\ref{friedr1}). We obtain:
\begin{equation}\label{eq:rhor}
 \rho_r(a)={\rho}_I\,\frac{D\,a^{4}}{\left[1+D\,a^{4}\right]^{2}}\,.
\end{equation}
In the previous formulas ${\rho}_I$ is the same quantity defined in
(\ref{eq:rhoI}). We confirm now that it represents the primeval energy
density for $a\to 0$, i.e. in the inflationary period, as it is
obvious from (\ref{eq:rLa}). Furthermore, since the (relativistic) matter energy
density $\rho_r\to 0$ as $a\to 0$, it follows that ultimately
$\rho_I$ is ``pure vacuum energy'' stored in the very early
Universe. As shown by the above equations, in the course of the fast
evolution the vacuum energy transforms into matter, so the Universe
becomes eventually dominated by the material component. The maximum
density of radiation achieved is $\rho_r^{\rm max}=\rho_I/4$. Beyond
that point the radiation density starts to decrease and at the same time the inflationary
period (which is powered only by the vacuum energy) comes naturally
to an end.  This is tantamount to saying that the $\bCDM$ model
incorporates natural inflation with ``graceful exit'', as we shall
further discuss below.

In the initial period, $D\,a^{4}\ll1$, the solution
\eqref{HS1} can be approximated by the constant value
(\ref{eq:deSitterSolution}), i.e. $H\simeq H_I$, the vacuum energy
density remains essentially constant and the scale factor increases
exponentially. This fact can be verified from (\ref{HS1}) by
integrating from some initial (unspecified) scale factor up to a
value $a(t)$ well within the inflationary period, in which
$D\,a^{4}\ll1$ still holds good. The result is just the inflationary
solution
\begin{equation}\label{eq:deSitter1}
a(t)=a_i\exp\left\{H_I t\right\}\,.
\end{equation}
The meaning of $t$ in this equation is the elapsed time within the
inflationary period, and  $a_i$ is the purported  value of the scale
factor at the time when inflation started (defined to be at $t=0$),
hence $a(0)=a_i$. What happened before that instant of time, we
don't care. We will probably never know since the pre-inflationary
Universe is not accessible.

Once the inflationary phase is left behind, the term $D\,a^{4}$ can
be comparable to $1$ or much greater. Here the integration of
Eq.\,(\ref{HS1}) must be done with the full expression, and we find
\begin{equation}\label{HS1b}
\int_{a_*}^a\frac{d\tilde{a}}{\tilde{a}}\left[1+D\,\tilde{a}^{\,4}\right]^{1/2}=
H_I\,t\,.
\end{equation}
In this case $t$ is (at variance with the previous case) the time
elapsed after approximately the end of the inflationary period,
indicated by $t_*$, and we have defined $a_*=a(t_*)$. The
integration constant $D$ is fixed from the condition $H(a_*)\equiv
H_*$, and it satisfies $Da_*^{4}=\left( H_I/{H_*}\right)^2-1$.

We can now show that the ``graceful exit'' from the inflationary
phase can be successfully accommodated in this class of
models\,\cite{H4model}.
In fact, considering the limit $D\,a^{4}\gg1$
of equation (\ref{HS1b}) we find
\begin{equation}
a\sim t^{1/2}\,,
\end{equation}
which confirms our contention that the vacuum phase decays into a
radiation-dominated Universe. We can reach the same conclusion from
the analysis of the energy densities in the transition period. In
effect, from (\ref{eq:rLa}) and (\ref{eq:rhor}) we see that in the
limit $D\,a^{4}\gg1$ the decay law of matter is the characteristic
one of radiation density, $\rho_r\sim a^{-4}$, whereas the vacuum
energy density decays much faster,  $\rL\sim a^{-8}$, so it becomes
suppressed as $\rL/\rho_r\sim a^{-4}$ once the radiation regime has
been settled. Such suppression is welcome, of course, as  the success of BBN (Big Bang Nucleosynthesis) is not
jeopardized in this model. Recall also that when $\nu\neq 0$, we have $|\nu|\ll1$
(cf. Sect. 7) so the $H^2$-term is also harmless at the BBN epoch. %

\begin{figure}
\begin{center}
\includegraphics[scale=0.55]{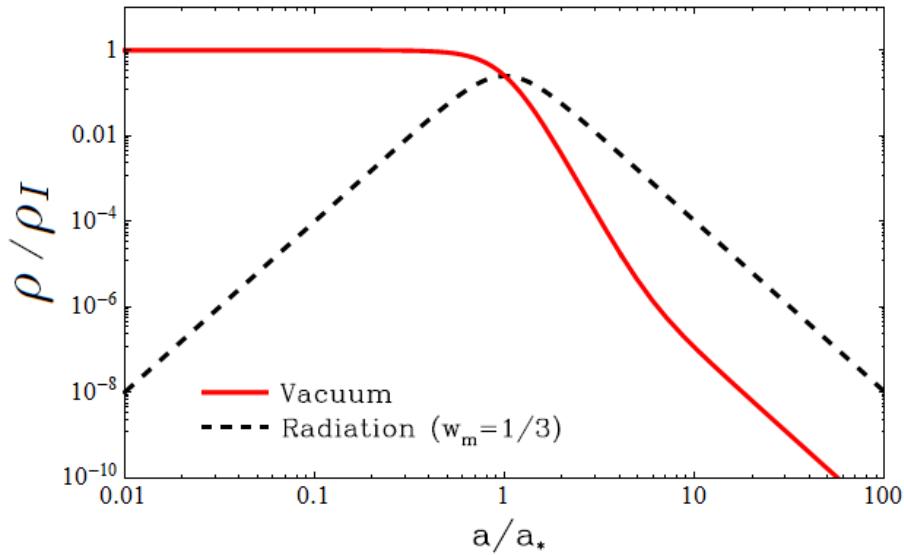}
  \caption{\scriptsize The transition from the
inflationary period of the running vacuum model (\ref{lambdaH2H4})
into the FLRW radiation epoch. The vacuum and matter energy
densities are normalized with respect to primeval density $\rho_I$,
and the scale factor is normalized with respect to the transition
one $a_{*}$ (see the text). The figure shows an initial (and
constant) dominance of the vacuum phase, and its subsequent decay
into relativistic particles (whose energy density rapidly increases)
until the Universe is  dominated by radiation and the vacuum energy
becomes negligible. As a result inflation ends with ``gracefully
exits'' into the standard radiation phase. We have used $8\pi G=1$
units.}
\label{Fig:VacuumDecay}
\end{center}
\end{figure}

As a concrete illustration, the evolution of the primeval matter and
vacuum energy densities from the very early Universe to the present
day is highlighted in Fig.\,\ref{Fig:VacuumDecay}, where we can
clearly see the transition from the de Sitter vacuum-dominated phase
into the radiation dominated phase.

\section{The horizon and entropy problems in the $\bCDM$}\label{sect:SolvingEntropy}

From the analysis of the previous section it follows that the
$\bCC$CDM Universe  (\ref{lambdaH2H4}) starts with a huge vacuum
energy density of the order $\sim M_X^4$ -- confer Eqs.\,(\ref{eq:rhoI})
and (\ref{eq:MXvalue}) -- and therefore without an initial
singularity, and has no horizon problem. It follows that a light pulse beginning in
the remote past at $t=t_1\gtrsim t_i$ (i.e. shortly after inflation
started at $t_i$) will have traveled until the end of inflation,
$t_f$, the physical distance
\begin{equation}\label{eq:horizon}
d_{H}(t_f)= a(t_f)\int_{t_1}^{t_f}\frac{d
t'}{a(t')}=\frac{a_f}{a_i\,H_I}\left(e^{-H_It_1}-e^{-H_It_f}\right)\simeq\,\frac{a_f\,e^{-H_It_1}}{a_iH_I}\,,
\end{equation}
where $a(t)$ is given by Eq.\,(\ref{eq:deSitter1}) during the
inflationary stage, and $H=H_I$ remains constant under inflation. In
practice $H$ need not be strictly constant, it is enough that it
remains approximately so for some interval $t_i<t_1<t_f$.  The
integral (\ref{eq:horizon}) diverges for $t_1\to t_i$ if the initial
time of inflation is a sufficiently early one (viz. if
$a_i\to 0$). This proves the non-existence of particle horizons
in the running vacuum Universe. Equivalently, and perhaps more
transparently, the above integral can be written as
\begin{equation}\label{eq:horizonInfinite}
d_{H}(a_f)= a(t_f)\int_{a_1}^{a_f}\frac{d
a'}{a'^2\,H}=\frac{a_f}{H_I}\left(\frac{1}{a_1}-\frac{1}{a_f}\right)\simeq
\left(\frac{a_f}{a_1}\right)\,H_I^{-1}\,.
\end{equation}
Being $a_1$ at time $t_1\gtrsim t_i$ exponentially smaller than
$a_f$ at the end of inflation, we have $a_f/a_1\ggg 1$ and so the
above integral (hence the horizon) can be as big as desired. The
result (\ref{eq:horizonInfinite}) is of course the same as
(\ref{eq:horizon}) because $a_1=a_i\,e^{H_It_1}$.

Notice that $\chi\sim d_H/a_f\simeq H_I^{-1}/a_1$ is the angle
subtending the horizon from the point where inflation starts. This angle is a comoving
coordinate, and if it is fixed appropriately at time $t_1$ (i.e. if
it is made large enough such that it subtends at least the
current horizon) it will remain so for the entire cosmic
history\,\footnote{For this it suffices that $a_f/a_1$ be only $\sim
1\%$ of the total expansion $a_0/a_f\sim 10^{28}$ after inflation.
Equivalently, the number of inflationary e-folds ($e^{N}=a_f/a_1$)
must be $N>26\ln 10\simeq 60$.}. The exponential solution triggered
by a period of approximately constant vacuum energy (starting from an
early point $a_1$) makes this arrangement possible and
hence there is effectively no horizon in the $\bCDM$ model.

In contrast, in the concordance $\CC$CDM model the above integral
gives (e.g. in the radiation dominated epoch) $d_H\sim a^2$  (this
is again $d_H\sim H^{-1}$, but now $H$ is not constant) and as a
result $d_H/a\to 0$ for $a\to 0$, which is tantamount to say that
the observers become isolated in the past (namely, their horizon
becomes infinitely smaller than the size of the Universe for $t\to
0$ and hence cannot exchange information with it) . This is the well-known horizon problem of the concordance model, which as we
have seen can be overcome in the $\bCDM$ model. On more physical
grounds, we can say that in the $\bCDM$ model the local interactions
(the exchange of information) are able to homogenize the whole
Universe since the very beginning.

\subsection{The large cosmological entropy}\label{sect:EntropyProblem}

We can reconfirm the absence of the horizon problem in the running
vacuum model from a thermodynamical perspective, namely from the
point of view of the so-called ``entropy
problem'' in cosmology. Succinctly formulated reads as follows: how did the universe manage to start with an extremely low entropy
value and develope later a huge one that gave rise to the arrow of time in accordance with the second law of thermodynamics? There is no solution to it within the $\CC$CDM.

Let us start by recalling some basic thermodynamical facts of our Universe that are intimately
connected with the horizon problem. The current horizon is defined
by  $H_0^{-1}$ and we normalize the scale factor as $a_0=1$. We can
take $H_0^{-3}$ as a fiducial volume of the observable Universe (up
to an irrelevant numerical factor common to all our estimates). From
standard thermodynamical formulae of the thermal
history\,\cite{BookKolbTurner} we can estimate the total entropy of
our Universe multiplying the entropy density by its fiducial volume.
In natural units, we find:
\begin{equation}\label{eq:TotalEntropyToday}
S_{0}=
\frac{2\pi^2}{45}\,g_{s,0}\,\Tgz^3\,\left(H_0^{-1}\right)^3\simeq
2.3 h^{-3} 10^{87}\sim 10^{88}\,.
\end{equation}
Here $\Tgz\simeq 2.725$$^{\circ}$K $\simeq 2.35\times 10^{-13}$ GeV
is the present CMB temperature, $H_0=2.133\,h\times 10^{-42}$ GeV
(with $h\simeq 0.67$) is the corresponding value of the Hubble function\,\cite{CosmoObservations2}, and the
coefficient $g_{s,0}=2+6\times
(7/8)\left(T_{\nu,0}/\Tgz\right)^3\simeq 3.91$ is the entropy factor
for the light d.o.f. today, which involves the well-known ratio
$T_{\nu,0}/\Tgz=\left(4/11\right)^{1/3}$ of the current neutrino and
photon temperatures.

However, in the context of the $\CC$CDM concordance model, the huge
number (\ref{eq:TotalEntropyToday}) cannot be understood without
generating a phenomenal causality problem, and hence recreating
the horizon problem again. The reason is that in the
$\CC$CDM  the total entropy contained in the horizon at earlier
times is much smaller. In fact, we can project the result
(\ref{eq:TotalEntropyToday}) backwards in time taking into account
that it evolves as $\left(T\,H^{-1}\right)^{3}$. Now, for an
adiabatic evolution $T\sim a^{-1}$. Moreover during the
nonrelativistic epoch $H\sim a^{-3/2}$, so that altogether this
renders $S\sim a^{3/2}= (1+z)^{-3/2}$. We conclude that at
recombination ($z_{\rm rec}\simeq 1100$) the total entropy was a
factor $\sim z_{\rm rec}^{-3/2}\sim 10^{-5}$ smaller, or, to be more
precise: $S_{\rm rec}\sim 10^{83}$. It means that the current Hubble
volume should contain some hundred thousand causally disconnected
regions at recombination! This is completely unacceptable since the
smoothness of the CMB must have a causal explanation in terms of
interactions that propagate at subluminal velocities. Such an
unsettling situation worsens more and more when we travel deeper and
deeper into the past, where the number of causally disconnected
regions keeps increasing inordinately. It goes without saying that
this a serious drawback of the standard picture, in fact one of the
biggest cosmological conundrums of the $\CC$CDM model!

\subsection{Solving the entropy  problem with running vacuum in a generic GUT}\label{sect:SolvingEntropyProblem}

The situation in the running vacuum Universe is completely
different. We shall show that the total entropy contained in the
current horizon $H_0^{-1}$ can be computed from $S_0\sim
S_{\infty}\,H_0^{-3}$, where $S_{\infty}$ is an asymptotic value
attained by the entropy per comoving volume (or ``comoving entropy''
for short) in the early Universe. Specifically we find that
$S(a)\sim T^3\,a^3\to S_{\infty}$ for $a\gg a_{eq}$, where $a_{eq}$
is the \textit{vacuum-radiation equality point}. This point is
placed in the (still incipient) radiation epoch, and in it the energy
density equals that of the decaying vacuum. We will show that the comoving
(hence intrinsic, volume-independent) radiation entropy increases
very fast ($S\propto a^6$) owing to the primeval vacuum decay until reaching that
saturation value $S_{\infty}$. It follows that the amount of entropy
contained in a physical patch of volume $V\sim H^{-3}$ (taken at any
$a\gg a_{eq}$) increases just proportional to the fixed number
$S_{\infty}$ times the physical volume of the region, i.e. $S\sim
S_{\infty}\,V$. In particular, for the region encompassed by the
current horizon, $V_0\sim H_0^{-3}$, we have $S_0\sim
S_{\infty}\,H_0^{-3}$. Remarkably, this value is
of the correct order of magnitude of the entropy today,
Eq.\,(\ref{eq:TotalEntropyToday}). Since the comoving entropy
$S/V\sim S_{\infty}$ is conserved after we have left well behind the vacuum-radiation equality point, there is no causality problem at all in the running $\bCDM$ model.

To substantiate these claims, let us first estimate the
equality point $a_{eq}$. It can be obtained from equating  the
densities (\ref{eq:rLa}) and (\ref{eq:rhor}), i.e.
$\rho_r(a_{eq})=\rL(a_{eq})$, hence $Da_{eq}^4=1$. This determines
$D=a_{eq}^{-4}$ and from (\ref{HS1}) we see that at $a=a_{eq}$ the
Hubble function reads $H_{eq}=H_I/\sqrt{2}$ (obviously below $H_I$).
As we shall see, the main results of this analysis do not depend on the details of the GUT framework that rules the dynamics of the vacuum. However, the precise value of  $a_{eq}$ is of course model dependent. We need not know it accurately, but it is convenient to have a numerical estimate.  Inserting $H_{eq}=H_I/\sqrt{2}$ and
$a=a_{eq}$ in Friedmann's equation relating the evolution of
radiation from the time when $H=H_{eq}$ up to now, yields
\begin{equation}\label{eq:aeqEXACT}
a_{eq}\simeq \left(\frac{8\pi^3\,g_{*}}{45}\right)^{1/4}\frac{\Tgz}{\sqrt{M_P
H_I}}=\left(\frac{\pi^2\,g_{*}}{15}\right)^{1/4}\frac{\Tgz}{M_X}\,,
\end{equation}
where in the last step use has been made of Eq.\,(\ref{eq:HIvalue}).
This result is of course not exact (as one cannot reach the equality point assuming that the evolution is always within the radiation epoch), but it provides the estimate $a_{eq}\sim 3\times 10^{-29}$, which is sufficient as an order of magnitude value\,\footnote{The early equality point $a_{eq}$ is specific of the $\bCDM$ Universe, and it should not be confused
with the standard equality point at which the energy densities of
radiation and cold dark matter become equal: $a_{EQ}\simeq 3\times
10^{-4}$ (i.e. at redshift $z_{EQ}\simeq 3300$). The latter
occurs much later and is coincident in both models. We have, $a_{EQ}\sim 10^{25} a_{eq}$!} for a typical GUT defined at the scale $M_X\sim 10^{16}$ GeV.

To compute the entropy generated from the primeval vacuum decay we
have to estimate the temperature of the radiation heat bath that is
formed during the conversion of the vacuum energy into
relativistic particles\footnote{See\,\cite{ThermalPaper2014} and \cite{CliftonBarrow2014} for recent alternative formulations, and \cite{Lima1996} for early developments.}. This post-vacuum heat bath will mimic the reheating process of the standard inflationary scenarios, but the process to achieve it is actually quite different. Let us call the radiation
temperature of the heat bath $T_r$.  Equating the radiation
density (\ref{eq:rhor}) emerging from vacuum decay in our model to
the general form of the radiation energy at an
effective temperature $T_{r}(a)$ we obtain:
\begin{equation}\label{eq:reheatingMatch}
\rho_r(a)={\rho}_I \frac{\left(a/a_{eq}\right)^4}{\left[1 +
(a/a_{eq})^{4}\right]^{2}}=\frac{\pi^2}{30} g_{\ast}
T_{r}^{4}(a)\,.
\end{equation}
Here  $g_{*}$ counts the total number of effectively massless
degrees of freedom (d.o.f.) in the heat bath at the given
temperature. For the standard model (SM) of particle physics,
$g_{*}=106.75$, if we include the top quark and the Higgs boson
since all of them can be relativistic d.o.f. at high
temperatures. In general $g_{*}$ will be larger than the SM value
when we operate at a GUT scale.

From Eq.\,(\ref{eq:reheatingMatch}) we find
$T_r$ as a function of the scale factor:
\begin{equation}\label{eq:Tr}
T_r(a)=T_X\,\frac{a/a_{eq}}{\left[1 +
(a/a_{eq})^{4}\right]^{1/2}}\,.
\end{equation}
Here
\begin{equation}\label{Temp}
T_X= \left(\frac{30\,\rho_I}{\pi^{2}g_*}\right)^{1/4} =
\left(\frac{30}{\pi^{2}g_*}\right)^{1/4}\,M_X
\end{equation}
is the value of the  temperature associated to the GUT scale $M_X$
through Eq.\,(\ref{eq:rhoI}). The temperature at $a=0$ is $T=0$. It then raises (linearly with $a$ in the beginning) until a maximum value, which is attained at $a=a_{eq}$. It is natural to think of such maximum value as a kind of effective ``reheating temperature'' after vacuum decay\,\footnote{Strictly speaking there is no ``reheating'' in this context since the Universe is just heating up progressively from $a=0$ till the point $a=a_{eq}$, where it reaches the maximum temperature (\ref{eq:TRH}). Later on, when the evolution enters fully the radiation epoch, the temperature falls down standard as in (\ref{eq:Tasymptotic}).}:
\begin{equation}\label{eq:TRH}
T_{RH}\equiv\frac{T_X}{\,\sqrt{2}}=\left(\frac{15}{2\pi^{2}g_*}\right)^{1/4}\,M_X\,.
\end{equation}
We point out that, in the conventional inflaton scenario\,\cite{BookKolbTurner}, the reheating temperature of the
radiation heat bath is not just determined by the initial GUT
temperature $T_X$. A new parameter comes into play, the inflaton
width $\Gamma_{\phi}$, and the reheating
temperature is $T_{RH}\sim g_{*}^{-1/4}\sqrt{M_P\Gamma_{\phi}}\sim
T_X\sqrt{\Gamma_{\phi}/H_I}$, where  $H_I^2\sim g_{*}\,
T_X^4/M_P^2$.

Notice that neither $T_X$ (nor $T_{RH}$ of course) are very
sensitive to the precise value of $g_{*}$, even if the number of
d.o.f. changes by one order of magnitude, because (\ref{Temp})-(\ref{eq:TRH}) depend on the
quartic root of $g_{*}$. Therefore we will just take the SM value
$g_{*}=106.75$ as a fiducial estimate for our calculations. In this
case we find e.g. $T_X\simeq 0.41\,M_X$, which is smaller than $M_X$
but not far away from it. As the GUT scale  $M_X$ is some three
orders of magnitude below the Planck mass, $M_P\sim 10^{19}$ GeV, we
are well reassured that our working regime is perfectly compatible
with a semiclassical description of QFT in curved spacetime (which
is the precise context where the running $\bCDM$ cosmological
framework is formulated\,\cite{JSP-CCReview2013}).

At the maximum temperature (\ref{eq:TRH}) the radiation density is
also maximum, so at that point the Universe becomes maximally
populated of relativistic particles emerging form vacuum decay.
Eventually both  the radiation energy and the temperature decay (for $a\gg
a_{eq}$), respectively as
\begin{eqnarray}
\rho_r(a)&\sim &\rho_I\,\left(\frac{a_{eq}}{a}\right)^4\sim M_X^4\,\left(\frac{a_{eq}}{a}\right)^4\label{eq:rhoasymptotic}
\end{eqnarray}
and
\begin{eqnarray}
T_r(a)&\sim& T_X\,\frac{a_{eq}}{a}\sim
M_X\,\frac{a_{eq}}{a}\label{eq:Tasymptotic}\,.
\end{eqnarray}
It follows that, in the asymptotic adiabatic regime, the thermodynamic behavior of the running vacuum model boils
down to the standard form, namely  $\rho_r\sim 1/a^4$
and $T\sim 1/a$. Thus, the conventional BBN picture can proceed
normally since the changes affect only the inflationary epoch.

Having found an expression for the effective temperature of the
primeval radiation heat bath, the comoving radiation entropy of the
relativistic particles that now populate the Universe can be
estimated. Starting from
\begin{equation}
S_{r}(a)= \frac{\rho_{r}(a) + p_{r}(a)}{T_{r}(a)} \, a^3 =
\frac{4}{3} \frac{\rho_{r}(a)}{T_r(a)}\, a^3= \frac{2\pi^2}{45}
g_{*} T_{r}^{3}(a)\, a^3\,,\label{eq19}
\end{equation}
and then using (\ref{eq:Tr}) we arrive at:
\begin{equation}
S(r) = \frac{2\pi^2}{45}\, {g_{*}}\,\left(
T_{X}^{3}\,a_{eq}^{3}\right) f(r)\,. \label{eq:Sr}
\end{equation}
We have defined the following function of the ratio $r\equiv
a/a_{eq}$:
\begin{equation}\label{eq:deffr}
f(r)=\frac{r^6}{\left(1+r^4\right)^{3/2}}\,.
\end{equation}
 Since $f(0)=0$, $f(1)=2^{-3/2}\simeq 0.35$ and $f(r)\to 1 $ for $r\gg1$, it follows that the radiation entropy (\ref{eq:Sr}) evolves monotonically very fast ($S\propto a^6$ in the beginning) starting from zero, then goes through the equality point $a=a_{eq}$ (where is still increasing) and finally reaches the maximum value
\begin{equation}
S(a) \to S_{\infty}\equiv\frac{2\pi^2}{45}\,
g_{*}\,\left( T_{X}^{3}a_{eq}^{3}\right) \ \ \ \ \ ({\rm
for}\ a\gg a_{eq})\,,\label{eq:Sinfty}
\end{equation}
where it saturates. This substantiates our claim on the rapid increase of the comoving entropy until achieving an asymptotic value $S_{\infty}$.

But of course we still have to prove that we can use $S_{\infty}$ to
predict the total entropy contained in our present horizon, Eq.\,
(\ref{eq:TotalEntropyToday}). We are now in position to complete
this task. First of all let us rewrite the expression $T_{X}^{3}a_{eq}^{3}$ appearing in parenthesis in the asymptotic formula (\ref{eq:Sinfty}) in terms of the temperature, $T_{rad}$, and the scale factor, $a_{rad}$, deep in the radiation epoch, i.e. $a_{rad}\gg a_{eq}$. From Eq.\,(\ref{eq:Tr}) we find:
\begin{equation}\label{eq:Trad}
T_{rad}=T_X\,\frac{a_{rad}/a_{eq}}{\left[1 +
(a_{rad}/a_{eq})^{4}\right]^{1/2}}\simeq T_X\,\frac{a_{eq}}{a_{rad}}\,.
\end{equation}
It is important to realize that this approximation is valid for any value of $a_{rad}$ well after we have passed the equality point $a_{eq}$, i.e. when we have entered the adiabatic regime of the evolution. From (\ref{eq:Trad}) it follows that $T_X\,a_{eq}\simeq T_{rad}\, a_{rad}$ in very good approximation in this regime. Inserting the result in  Eq.\,(\ref{eq:Sinfty}), we find that the total entropy in the current physical horizon is predicted to be
\begin{equation}
S_{\infty}\,H_0^{-3}=\frac{2\pi^2}{45}\,
\left(g_{*}\, T_{rad}^{3}\,a_{rad}^{3}\right)\,H_0^{-3}=\frac{2\pi^2}{45}\,
\left(g_{s,0}\,\Tgz^{3}\,a_0^{3}\right)\,H_0^{-3}=S_0\,,\label{eq:SinfH}
\end{equation}
where in the last step we have used the entropy conservation law of the adiabatic regime, which states that $g_{*}\, T_{rad}^{3}\,a_{rad}^{3}= g_{s,0}\,\Tgz^{3}\,a_0^{3}$. Recall that in our normalization $a_0=1$, and hence we have reproduced the result $S_0$  given by (\ref{eq:TotalEntropyToday}).

The upshot is significant: it turns out that the theoretical prediction (\ref{eq:SinfH}) of the running $\bCDM$ model for the total entropy enclosed in our current horizon is precisely the observed entropy in
our Universe, i.e. the huge entropy number (\ref{eq:TotalEntropyToday}) -- in natural units. This result may be viewed as a  potential solution to the cosmological entropy
problem in the context of the running $\bCDM$. As we have seen, we can assume some generic GUT at the very high scale, which triggers the decay of the huge vacuum energy and generates the large entropy in the radiation phase. Remarkably, the important CMB constraint (\ref{eq:HIlimit}) is preserved and the final result for the entropy is universal, meaning that it does not depend on the details of the GUT. The universality of the prediction is a reflex of the vacuum decaying dynamics and of the entropy conservation law, which holds good in the Universe once the cosmic evolution has entered the adiabatic regime.

The vacuum model (\ref{lambdaH2H4}) that we have studied here (in which the highest power of the Hubble rate is $H^4$) is just the simplest implementation of the running $\bCDM$ as a candidate model for a complete description of the cosmic history from the inflationary times to the present day. However, the main results are maintained if an arbitrary even power  $H^{2n}$ ($n>2$) of the Hubble rate is used for the higher order term (as required by the covariance of the effective action\,\cite{JSP-CCReview2013}). At the same time, one has to keep the additive term so as to reproduce the standard $\CC$CDM model at low energies. In Sect.\,\ref{sect:DynamicalVacuumCurrentUniverse} we shall see that the observations are compatible with keeping also the $H^2$-term of (\ref{lambdaH2H4}). This is interesting as the latter endows the current cosmological vacuum with some mild dynamics (as a kind of  ``smoking gun'' of the entire vacuum decaying mechanism) and can be interpreted as dynamical dark energy.

\section{Starobinsky inflation versus running vacuum model}\label{sect:StarobinskyRunning}

In this section we wish to elaborate on the fact that the class of vacuum models (\ref{lambdaH2H4}) realize in an effective way the Starobinsky type of inflationary regime\,\cite{Starobinsky80}\footnote{For recent and
past detailed studies of Starobinsky inflation, see e.g.
\cite{Vilenkin1985,Copeland2013}, and references therein.}. This
occurs somehow through the dominance of the highest power $H^4$  in
the early Universe, which is of the order of the Starobinsky's
correction term $R^2\sim H^4$ in the effective action, see
Eq.\,(\ref{eq:StarobinskyAction}) below. In actual fact the
situation is a bit more complicated since at the level of equations
of motion the term $R^2$ in the action gives rise not just to a pure $H^4$ contribution but to a fairly involved combination of powers of $H$ and its time derivatives.
Still there are some interesting similarities worth noticing. Some
of them have been recently pinpointed
in\,\cite{Essay2014,SUGRA2014}, and explored in great detail in
\cite{GoSol2014}.

\subsection{The standard Starobinsky action}

Let us recall that the original Starobinsky
model\,\cite{Starobinsky80} is based on the following action:
\begin{equation}\label{eq:StarobinskyAction}
S=\int d^4x\sqrt{-g}\left(\frac{R}{16\pi G}+bR^2\right)+S_{\rm
matter}\,,
\end{equation}
where $b$ is the dimensionless coefficient of the higher order
derivative term $R^2$. It is usually written as $b=M_P^2/6M_s^2$,
where $M_s$ is a mass dimension parameter -- playing the role of
scalaron mass in the original model\,\cite{Starobinsky80}.  Recall
that $R$ is, if written in the context of the FLRW metric, a linear
combination of $H^2$ and $\dot{H}$, and therefore $R^2$ involves
terms of the form $H^4$, $\dot{H}^2$ and $H^2\,\dot{H}$, all of
which are roughly speaking of order $\sim H^4$, and hence one may
envision a kind of close connection of the Starobinsky model and the
$\bCDM$ model. This is true in part, but it is not quite so, at
least not for the original form (\ref{eq:StarobinskyAction}) of the
Starobinsky action. In the next section we briefly mention
how to improve the connection of the running $\bCDM$ with an
alternative Starobinsky-like action\,\cite{Fossil07}. At the moment
we want to continue with the action (\ref{eq:StarobinskyAction}) and
show the form of the inflationary solution that emerges from it, as
this will enable us to compare with the inflationary scenario
derived from the $\bCDM$ model in Sect.4.

The equations of motion for the case where the energy-momentum
tensor of matter contains a mixture of various fluids $N=1,2...$
with densities and pressures $(\rho_N,p_N)$ can be obtained after
varying the action (\ref{eq:StarobinskyAction}) with respect to the
metric. The result reads as follows:
\begin{eqnarray}\label{eq:FieldEquationsStaro}
G_{\mu\nu}+32\pi Gb(\nabla_\mu\nabla_\nu R&-&g_{\mu\nu}\Box R+RR_{\mu\nu}-\frac{g_{\mu\nu}}{4}R^2)\nonumber\\
&=&8\pi G\sum_{N}\left[p_{N}g_{\mu\nu}-(\rho_{N}+p_{N})U^N_\mu
U^N_\nu\right]\,.
\end{eqnarray}
For $b=0$ we recover, of course, the standard Einstein's equations.
However, for $b\neq 0$ the result is more complicated. Setting
$(\mu,\nu)=(0,0)$ and $(\mu,\nu)=(i,j)$ and using the flat FLRW
metric, we find two independent equations:
\begin{equation}\label{eq:FriedmannEq}
H^2=\frac{8\pi G}{3}\sum_N\rho_N+96\pi
Gb(\dot{H}^2-2H\ddot{H}-6H^2\dot{H})
\end{equation}
and
\begin{equation}\label{eq:PressioEq}
H^2+\frac{2\dot{H}}{3}+32\pi
Gb(2\dddot{H}+12\ddot{H}H+18H^2\dot{H}+9\dot{H}^2)=-\frac{8\pi
G}{3}\sum_N p_N.
\end{equation}
If we now just project the result for a single matter component
$(\rR,p_R)$, and assume that this component is relativistic
($p_R=\rR/3$), we can combine the two previous expressions to obtain
a single differential equation for the Hubble rate:
\begin{equation}\label{eq:HubbleEq}
2H^2+\dot{H}+48\pi
Gb(2\dddot{H}+14\ddot{H}H+24H^2\dot{H}+8\dot{H}^2)=0\,.
\end{equation}
It is pretty clear that this equation (a third order, nonlinear,
differential equation) is considerably more involved than the first
order equation (\ref{HE}) for the $\bCDM$ model in the early
Universe. This shows that the initial analogy between the two
models, based on the fact that $R^2\sim H^4$, was an exceeding
simplification since the models have indeed to be compared at the
level of the equations of motion.

\begin{figure}
\begin{center}
\includegraphics[scale=0.42]{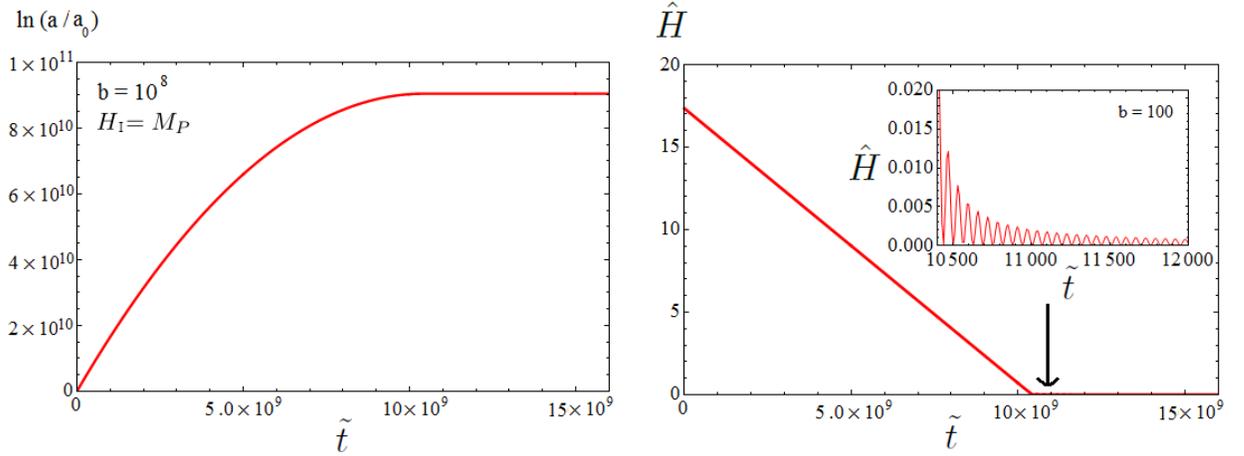}
  \caption{\scriptsize The inflationary solution of (\ref{eq:ScaleFactorEq}) corresponding to the Starobinsky model (\ref{eq:StarobinskyAction}).  On the left it is shown the exponential growth $a\sim e^{H_I\,t}$ of the scale factor and its stabilization into the radiation regime $a\sim t^{1/2}$. We have taken $H_I=M_P$ in (\ref{eq:CondicionsInicials}), and $b=10^8$ from the mentioned bound in the text. On the right we depict the corresponding behavior of the Hubble function and (in the inner window) the characteristic oscillations when the Universe leaves the inflationary phase and enters the radiation epoch in the form $a\sim t^{1/2}+{\it oscillations}$. In that window we have set $b=100$ to make the oscillations more clearly visible. Time has been rescaled as $t=(\sqrt{96\pi}/M_P)\,\hat{t}$, and $\hat{H}=(1/a)da/d\hat{t}$ is the correspondingly rescaled Hubble function. \label{Fig:Starobinsky}}
\end{center}
\end{figure}

Expressing (\ref{eq:HubbleEq}) in terms of the scale factor it
becomes a fourth order differential equation:

\begin{equation}\label{eq:ScaleFactorEq}
\frac{\ddddot{a}}{a}+3\frac{\dddot{a}\dot{a}}{a^2}+\frac{\ddot{a}^2}{a^2}-5\frac{\ddot{a}\dot{a}^2}{a^3}+\frac{M_p^2}{96\pi
b}\left(\frac{\dot{a}^2}{a^2}+\frac{\ddot{a}}{a}\right)=0\,.
\end{equation}
This equation is to be solved under initial conditions, which we
take as follows:
\begin{equation}\label{eq:CondicionsInicials}
a(0)=a_i\qquad \dot{a}(0)=a_i\,H_I\qquad \ddot{a}(0)=a_i\,H_I^2\qquad
\dddot{a}(0)=a_i\,H_I^3
\end{equation}
and assume $H_I\sim M_P$ in this case. An analytic solution of this
problem is not possible in general. However, for $b\to 0$ the
dominant term is the last one of  (\ref{eq:ScaleFactorEq}), and the
solution becomes $a\sim t^{1/2}$ corresponding to the radiation
epoch. In contrast, for large $b$ the last term of
(\ref{eq:ScaleFactorEq}) becomes negligible and, then, the solution
is easily seen to be the exponential one: $a(t)=a_i\,e^{H_I\,t}$.
The numerical solution of Eq.\,(\ref{eq:ScaleFactorEq}) under the
initial conditions (\ref{eq:CondicionsInicials}) is given in
Fig.\,\ref{Fig:Starobinsky}. Remarkably an approximate
solution of (\ref{eq:ScaleFactorEq}), or equivalently of
(\ref{eq:HubbleEq}), can be found to describe the end of the
inflationary period. Since $\dot{H}$ remains essentially constant
until we are very near the end of the inflationary phase (see the
straight line in the plot on the right in
Fig.\,\ref{Fig:Starobinsky}), we can solve (\ref{eq:HubbleEq}) by
neglecting $\dot{H}/H^2\ll1$ and all higher derivative terms. In
this way we are left with the equation $576\pi G_N\,b\,\dot{H}=-1$, whose
immediate solution is $H(t)=H_I-{M_P^2t}/{576\pi b}$, and hence the
corresponding scale factor reads:
\begin{equation}\label{eq:ScaleFactor1P}
a(t)=a_i\,e^{H_It}e^{-\frac{M_P^2t^2}{1152\pi b}}\,.
\end{equation}
Obviously, $b>0$ in order to have a stable inflationary solution until
the inflationary phase terminates at around $t_{f}\simeq
1152\,\pi\,b/M_P$. The larger is $b$ the longer is the
inflationary time. One can show that if we impose the limit
(\ref{eq:HIlimit}) from the CMB anisotropies the parameter $b$ has a
lower bound of order $10^8$ -- see\,\cite{Copeland2013}.

Needless to say the approximate solution
(\ref{eq:ScaleFactor1P}) does not perfectly interpolate from the
inflationary epoch to the radiation epoch and hence, in contrast to
the situation with the $\bCDM$ model, we do not have an analytically
rigorous description of the graceful exit. The solution
(\ref{eq:ScaleFactor1P}) is only indicative that the inflationary
process comes to an end, and that there is a chance for a correct
transition to the $a\sim t^{1/2}$ radiation epoch. But it misses the
nontrivial details accounting for the reheating stage. Unfortunately, the issue of graceful exit is still a complicated and unsolved matter in Starobinsky inflation,
despite it has been discussed in different places in the literature,
see e.g. \cite{Vilenkin1985} and
\cite{ShapSol2002} in different formulations. A more detailed study,
though, is still needed\,\cite{GoSol2014}.

In point of fact, we see from the comparison of Figures 1
and\,\ref{Fig:Starobinsky} that the connection of the Starobinsky
model and the $\bCDM$ model is not that close after all, at least
using the original action (\ref{eq:StarobinskyAction}). It turns out
that the connection with the $\bCDM$ model is much better
accomplished if one adopts the conformally invariant formulation of
the Starobinsky model, which is only broken by quantum effects; namely,
the so-called anomaly-induced effective action (see next section).
The reason is that only in this formulation the $\bCDM$ model can be
motivated from an effective action, which is certainly not the one
given by (\ref{eq:StarobinskyAction}). The latter is, in effect, not
conformally invariant.

\subsection{The anomaly-induced effective action}

A generalized form of Starobinsky's inflation, based on the
effective action of anomaly-induced inflation was considered long
ago in\,\cite{ShapSol2002,DESYTomsk2002,Takakura2003} --  and
references therein --  but the first explicit connection of this framework with the running vacuum model $\bCDM$ was signaled in \cite{Fossil07}. One finds that by taking the masses of the fields into account, using the
conformization procedure of the cosmon model\,\cite{PSW}, the
inflationary process automatically slows down and can therefore
favor the graceful exit.  Let us briefly mention the solution of
this alternative form of Starobinsky inflation. We follow
Ref.\,\cite{Fossil07}, where more details are provided. Rather than
starting from the action (\ref{eq:StarobinskyAction}) one takes the
classically conformally invariant higher derivative action of the
vacuum:
\begin{equation}\label{EAvacuum}
S_{\rm vac}\, =\, \int d^4x\sqrt{-g}\, \left\{ a_1 C^2 + a_2 E + a_3
{\nabla^2} R \right\}\,,
\end{equation}
where $a_{1,2,3}$ are dimensionless coefficients, $C^2$ is the
square of the Weyl tensor and $E$ is the Gauss-Bonet topological
invariant in $4$ dimensions\,\cite{BooksQFTcurved}. This action can
then be completed with the conformally invariant realization of the
Einstein-Hilbert (EH) term:
\begin{equation}\label{eq:EHc}
S_{EH}^c =\frac{M_P^2}{16\pi\,\cM^2}\,\int d^4 x\sqrt{-g}\,
\left[\,R\chi^2 - 6\,(\partial \chi)^2\,\right]\,.
\end{equation}

Here $\chi$ is a background scalar field that realizes the conformal
symmetry at the high energy scale $\cM$, presumably close to the GUT
scale $M_X$. The same field and scale are used to conformize the
masses of all the matter fields in the action\,\cite{Fossil07}
following the prescription of\,\cite{PSW}. After $\chi$ acquires a
vacuum expectation value of order $\cM$, that symmetry is
spontaneously broken and we recover the standard EH term. Now, of
course, the conformal symmetry of the total action is also
intrinsically broken by quantum effects, not only the traditional
ones from the trace anomaly associated to the $S_{\rm vac}$
part\,\cite{BooksQFTcurved} but also by the additional contribution
form $S_{EH}^c$. The anomalous induced part of the action (i.e. that
one breaking conformal symmetry from quantum effects) follows from
solving the functional differential equation:
\begin{eqnarray}\label{eq:TraceAnomaly}
\lefteqn{ {<T_\mu^\mu> = -\,\frac{2}{\sqrt{-g}} g_{\mu\nu}
\,\frac{\delta \bar{\Gamma}}{\delta g_{\mu\nu}} +
\frac{1}{\sqrt{-g}}\,\chi\, \frac{\delta\bar{\Gamma}}{\delta
\chi}}}\mbox{\hskip 8cm}\\
\lefteqn{ {= b_1C^2\,+\,b_2E\,+\,b_3{\nabla^2}R
\,+\,\frac{\tilde{f}}{16\pi G_N\cMd} \,[R\chi^2-6(\partial \chi)^2]
\,.}}\mbox{\hskip 8cm}\nonumber
\end{eqnarray}
The one-loop values of the $\beta$-functions $b_1,b_2,b_3$ are well
established since long time ago\,\cite{BooksQFTcurved} and depend on
the matter content of the model. The new ingredient is the
$\beta$-function of the conformal EH term, which is given
by\,\cite{Fossil07}
\begin{equation}\label{ftilde}
\tilde{f} =
\frac{1}{3\pi}\sum_{F}\,\frac{N_F\,\,m_F^2}{M_P^2}+\frac{1}{2\pi}\sum_{V}\,\frac{N_V\,\,M_V^2}{M_P^2}\,.
\end{equation}
It involves contributions from fermions (F) and vector bosons (V)
since the scalar part vanishes in the classical conformal limit. The
trace-anomaly equation (\ref{eq:TraceAnomaly}) can be solved
following  the standard method\,\cite{AnomInEA}, namely from a
conformal transformation of the metric
$g_{\mu\nu}=e^{2\sigma}\bar{g}_{\mu\nu}$,  extended with the new
background field: $\chi=e^{-\sigma}\bar{\chi}$.  In this way one
finds the corresponding effective action that includes the effects
of $S_{EH}^c$ \,\cite{ShapSol2002,Fossil07}. The relevant
part of it can be summarized as follows:
\begin{eqnarray}\label{Seff2}
S_{\rm eff}= \int d^4 x \sqrt{-{\bar g}}\,
\frac{\bar{M}_P^2(\sigma)}{16\pi{\cal M}^2}
\left[\bar{R}\bar{\chi}^2- 6\,(\partial\bar{\chi})^2\right]
+S_{matter}+\,{\rm high.\ deriv.\ terms}\,,
\end{eqnarray}
where
\begin{equation}\label{eq:runningPlanckMass}
\bar{M}_P^2(\sigma)=M_P^2(1-\tilde{f}\,\sigma)
\end{equation}
is a ``running Planck mass'' that is sensitive to the dynamics of the
conformal factor $\sigma$.
We may next fix the conformal gauge as $\chi={\cal M}$ (i.e.
$\bar{\chi}={\cal M}e^{\sigma}$) so as to recover the EH form of the
action. The scale ${\cal M}$ cancels and the resulting dynamics does
not depend on it. Finally, we know that the FLRW metric is
conformally flat in conformal time $\eta$, so that with the  choice
$e^{2\sigma}=a^2(\eta)$ we can take the flat metric
$\bar{g}_{\mu\nu}=\eta_{\mu\nu}$. This renders the new curvature
terms trivial, $\bar{R}=0$, but of course the action depends in a
nontrivial way on the conformal factor $\sigma=\ln a(\eta)$. Varying
this action with respect to it and reverting to the cosmic time $t$
one finds a fourth order differential equation for $a(t)$ similar to
(\ref{eq:ScaleFactorEq}), although not identical. We shall omit
cumbersome details here\,\cite{ShapSol2002,Fossil07}, but we remark
that an approximate solution of it can also be found in the form
(\ref{eq:ScaleFactor1P}) as follows:
\begin{equation}\label{TAIS}
a(t)=a_i\,e^{H_I t}\,e^{-\frac14 H_I^2\tilde{f} t^2}\,.
\end{equation}
We see that $\tilde{f}$ plays, up to a factor, the role of
$1/b=6M_s^2/M_P^2\,$ in the original model, and so the effective
scalaron mass squared $M_s^2$ is substituted here by a combination
of squared masses of fermion and boson fields -- confer
Eq.\,(\ref{ftilde}).

With the anomaly-induced formulation we have recovered once more a
``tempered'' form of inflation, namely possessing an evolution pattern
very similar to that indicated in Fig.\,\ref{Fig:Starobinsky}, and thus
presumably having a paved way to graceful exit. Not
only so, now with the alternative formulation we have gained an
additional bonus, to wit: the running Planck mass squared
(\ref{eq:runningPlanckMass}) leads to a an effective gravitational coupling
\begin{equation}\label{Ga}
\bar{G}_N(a)=\frac{G_N}{1-\f\ln a}\,,
\end{equation}
where $G_N=1/M_P^2$ is the current value ($a=1$). Let us assume that
this situation is roughly maintained at low energies\,\footnote{We
cannot exclude that as we run down to the low energy regime,
corresponding to the present Universe, there might occur some
additional, infrared, renormalization of the coupling $\tilde{f}$.
This could enhance its value as compared to the UV regime.}. If we
next insert (\ref{Ga}) into the general low-energy equation for
energy conservation in the presence of variable cosmological
parameters $G_N$ and $\rL$, i.e. Eq.\,(\ref{BianchiGeneral}), we
obtain a differential equation relating $\rL$ and $\rmr$. However,
if we further assume matter conservation, then we can derive the
evolution law for the vacuum energy density $\rL$ that is compatible
with (\ref{Ga}) according to the Bianchi identity (\ref{Bianchi1}).
The exact result is a bit complicated, but in the
limit $|\tilde{f}|\ll 1$ (which is in fact the natural situation for
a $\beta$-function coefficient) it turns out that the final result
for the matter-dominated epoch reads very simple\,\cite{Fossil07}:
\begin{equation}\label{rLaAnomInd}
\rL(H)=\rLo +\frac{\tilde{f}}{8\pi\,G}\left(H^2-H_0^2\right)\,,
\end{equation}
where $\rL(H_0)=\rLo$ is the current value. This expression is
noticeable as it shows that the vacuum energy of the anomaly-induced
formulation of the Starobinsky inflation leads, at low energies, to
a dynamical vacuum model of type A1, see Eq.\,(\ref{lambdaTypeA1}).
We can easily identify by comparison of the two equations that
$\nu=\tilde{f}/3$.  It shows that the coefficient $\nu$ (which could have been introduced in (\ref{lambdaTypeA1}) on mere phenomenological grounds) can be interpreted as a $\beta$-function coefficient for the CC running in semiclassical curved spacetime. One can indeed verify that the structure of $\tilde{f}$ in (\ref{ftilde}) takes on the expected general form (\ref{eq:nualphaloopcoeff}). This result reinforces the physical meaning of the general renormalization group equation of the cosmological term, Eq.\,(\ref{seriesLambda}), in QFT in curved spacetime.

At the end of the day, we conclude that the class of vacuum models
(\ref{lambdaH2H4}) encodes basic features of different
realizations of Starobinsky's type of inflation, mainly if
realized in the anomaly-induced form. It remains to be seen what is
the final observational status of the Starobinsky model. At the day
of writing this work the situation is
inconclusive\,\cite{BICEP2a,PlanckRevision} and therefore we must
wait for more input.

\begin{figure}
\begin{center}
\includegraphics[scale=0.55]{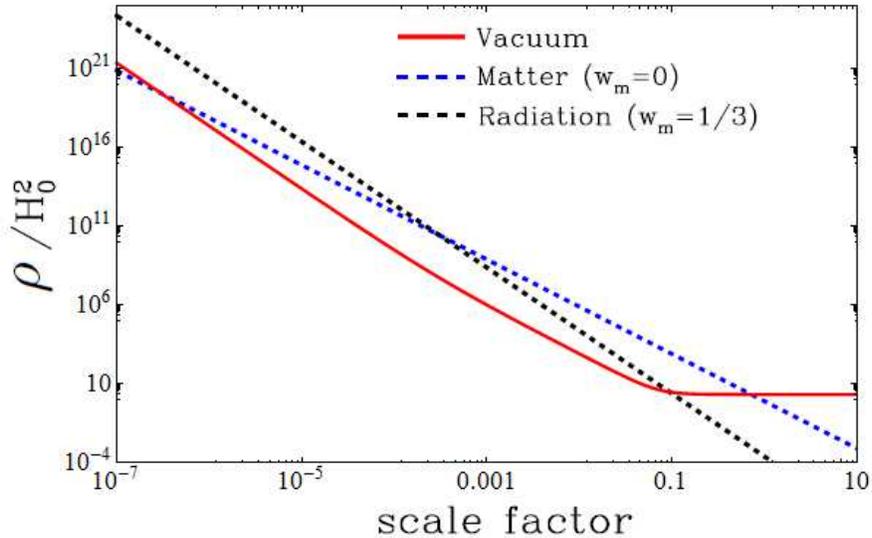}
\caption{\scriptsize The evolution of the energy densities (normalized with
respect to $H_0^2$) for the type-A1 model (\ref{lambdaTypeA1}) in
the post-inflationary epoch until our days. The curves shown are:
radiation (dashed line), non-relativistic matter (dotted line) and
vacuum (solid line, in red). Used inputs: $\nu=10^{-3}$,
$\Omega^{0}_{m}=0.27$,
$\Omega^{0}_{R}=(1+0.227N_{\nu})\,\Omega^{0}_{\gamma}$,
$(N_{v},\Omega^{0}_{\gamma},h)\simeq (3.04,2.47\times
10^{-5}h^{-2},0.71)$.  We have used $8\pi G=1$
units.} \label{Densities}
\end{center}
\end{figure}

\section{Vacuum dynamics from $\bCDM$ in the current Universe}
\label{sect:DynamicalVacuumCurrentUniverse}

In this section we go phenomenological and report on the
confrontation of the dynamical vacuum models A1,A2, B1 and B2
defined in Sect. 3 against the latest observational data. All of
these  models can be solved both at the background and perturbative
level. The Type-B ones, however, are more difficult to deal with
analytically and require more numerical work, specially in regard to the
perturbation equations and the study of structure
formation.  For details see the recent works\,\cite{GoSolBas2014,SolGo2014},
which are to our knowlwdge the most comprehensive studies on these matters currently existing in the literature. Here we limit ourselves to
provide a quick summarized presentation of the solution of models A1
and A2 only, but we provide also some numerical results of the type-B ones.

\subsection{The cosmology of type-A models}
\label{sect:CosmoTypeA}

Let us consider the vacuum model A2 defined in
Eq.\,(\ref{lambdaTypeA2}). Using the equations of Sect. 2 we can
solve for the energy densities as a function of the scale factor.
For cold matter and radiation one finds:
\begin{equation}\label{MatterdensityCCtCDM}
\rho_m(a) =  \rmo~a^{-3 \xi}\,,\ \ \ \ \  \rho_r(a) = \rho_r^0
~a^{-4 \xiR}\,
\end{equation}
and for the vacuum energy density:
\begin{equation}\label{rLArad}
\rL(a)=\rLo+{\rmo}\,\,(\xi^{-1} - 1) \left( a^{-3\xi} -1  \right) +
{\rRo}\,\,(\xiR^{-1} - 1) \left( a^{-4\xiR} -1\right)\,,
\end{equation}
where we have defined
\begin{equation}\label{defxiM}
\xi= \frac{ 1 - \nu }{ 1 - 3\tilde{\nu}/2 }\,, \ \ \ \ \ \ \ \ \
\xiR= \frac{ 1 - \nu }{ 1 - 2\tilde{\nu}}\,.
\end{equation}
Obviously, for $\tilde{\nu}=0$ the model A2 becomes simply the basic
running vacuum model A1 (introduced in \cite{SS2000}). The
corresponding Hubble function reads
\begin{equation}\label{HArad}
 H^2(a)=H_0^2\left[1+\frac{\Omo}{\xi}\left(a^{-3\xi}-1\right)+\frac{\Oro}{\xiR}\left(a^{-4\xiR}-1\right)\right]\,,
\end{equation}
where the cosmological parameters satisfy the usual sum rule $\Omo +
\Oro+\OLo = 1$.
As expected, for $\nu=\tilde{\nu}=0$ (i.e. $\xi=\xiR=1$) the Hubble
function $H(a)$ takes on the form of the concordance model. In this
case we recover $\rmr\sim a^{-3}$, $\rho_r\sim a^{-4}$ and, of
course, also $\rL=$const. As an illustration, in Fig.
\ref{Densities} the evolution of the matter and vacuum densities are
plotted from the radiation epoch to our days. It corresponds to a
type-A1 model ($\tilde{\nu}=0$) and we have assumed $\nu=10^{-3}$,
i.e. a typical value obtained from the fits to the observational
data, see below. We can see that the vacuum density performs a
substantial evolution until it asymptotes to the current value
$\rL\to\rLo$. It certainly looks more natural than the situation in
the $\CC$CDM model, where it stays $\rL=\rLo$ all the time since the
radiation epoch.

\begin{table*}
\begin{center}
{\begin{tabular}{| c |  c | c | c | c | } \multicolumn{1}{c}{Model}
& \multicolumn{1}{c}{$\Omo$} & \multicolumn{1}{c}{$\nu$} &
\multicolumn{1}{c}{$\epsilon$} & \multicolumn{1}{c}{$\chi^2/dof$}
\\ \hline
$\CC$CDM & $0.293\pm 0.013$ & - & - & $567.8/586$ \\\hline $A1$ & $0.292\pm 0.014$ & $+0.0013\pm 0.0018 $ & - &
$566.3/585$   \\\hline $A2$ & $0.290 \pm 0.014$ & $+0.0024\pm
0.0024 $ & - & $565.6/585$  \\\hline $B1$ & $
0.297^{+0.015}_{-0.014}$ & -  & $-0.014^{+0.016}_{-0.013}$ &
$587.2/585$   \\\hline $B2$ & $ 0.300^{+0.017}_{-0.003}$ &$ -
0.0039^{+0.0020}_{-0.0021}$  & $- 0.0039^{+0.0020}_{-0.0021}$ &
$583.1/585$   \\\hline
 \end{tabular}}
\caption{\scriptsize The fit values for the various models using SNIa+CMB+BAO$_{dz}$
data, together with their statistical significance according to
$\chi^2$ statistical test. For model A2 we provide the fit value of
$\nueff$ (see the text). For model B2, we have set $\nu=\epsilon$. \label{tableFitBAOdz}}
\end{center}
\end{table*}

To study the evolution of matter perturbations we generalize the
standard second order differential equation\,\cite{Peebles1993} for the growth factor
$D\equiv\delta\rmr/\rmr$ for the case when the vacuum term is
dynamical, $\dot{\rho}_{\CC}\neq 0$. The result (after neglecting
contributions which are subleading) is\,\cite{GoSolBas2014}:
\begin{equation}\label{diffeqD}
\ddot{D}+\left(2H+\Psi\right)\,\dot{D}-\left(4\pi
G\rmr-2H\Psi-\dot{\Psi}\right)\,D=0\,,
\end{equation}
with $\Psi\equiv -{\dot{\rho}_{\CC}}/{\rmr}$. For the $\CC$CDM model
we have $\rL=$const. and hence $\Psi=0$, so that the above equation
correctly reduces to the standard one\,\cite{Peebles1993}.

\subsection{Confronting the vacuum models A and B with observations}
\label{sect:ConfrontingTypeATypeB}

Equation (\ref{diffeqD}) can be solved for type-A models (for which
$\rmr$ and $\rL$ have been given above) in terms of hypergeometric
functions. In the case of type-B models the solution is more
complicated and one has to use direct numerical
methods\,\cite{GoSolBas2014,SolGo2014}. From the corresponding solution for
$D(a)$ we can e.g. study the linear growth rate of clustering\,\cite{Peebles1993},
namely the logarithmic derivative of the linear growth factor $D(a)$
with respect to $\ln a$:
\begin{equation}\label{eq:growingfactor}
f(a)\equiv \frac{1}{D(a)}\frac{dD(a)}{d\ln a}=\frac{d\ln D(a)}{d{\rm
ln}a}=-(1+z) \frac{d\ln D(z)}{dz}\,.
\end{equation}
This quantity is important enough since it is measured
observationally, so we can use it to investigate the performance of
our vacuum models. In Fig.\,\ref{GrowingRate} we compare the
theoretical growth prediction derived from our fits to the data (see
below) with the latest growth data (as collected e.g. by
\cite{BasilakosNes2013} and references therein). Specifically we
plot the combined observable $f(z)\sigma_{8}(z)$, viz. the ordinary
linear growth rate weighted by the rms mass fluctuation field, which
it is claimed to have some advantages in practice\,\cite{Song09}.
The value of $\sigma_{8}(z)$ can also be computed for each model
through $\sigma_8(z)=\sigma_8\,{D(z)}/{D(0)}$, where we use
$\sigma_8=0.829\pm 0.012$ from Planck+WP\,\cite{CosmoObservations2}.


\begin{figure}[!t]
\begin{center}
\includegraphics[scale=0.35]{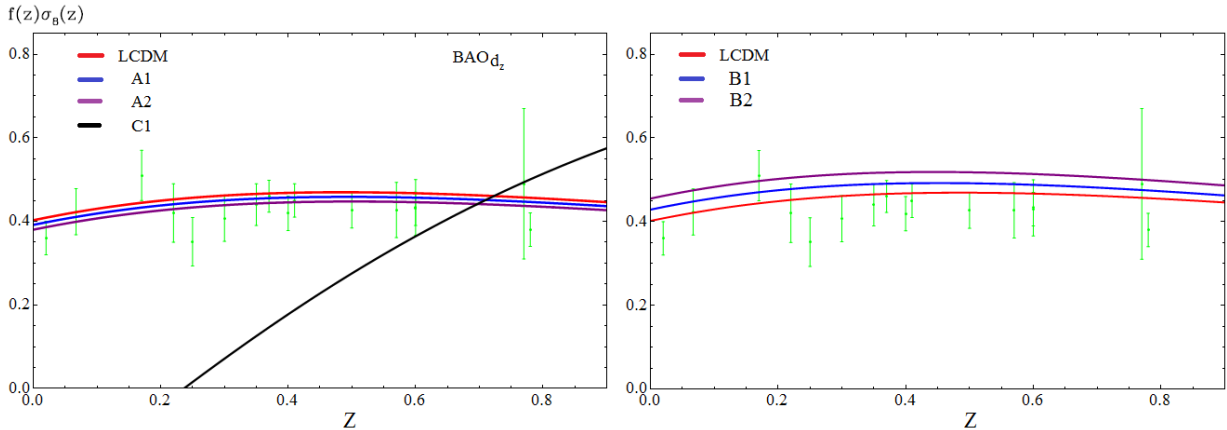}
\caption{\scriptsize Comparison of the observed (solid points
with vertical error bars) and theoretical evolution of the weighted
growth rate $f(z)\sigma_{8}(z)$ for the various A and B models. The
uppermost (red) line corresponds to the $\CC$CDM model, used as a
reference. The curve C1 (black line) that deviates significantly
from the others in the two panels corresponds to model B2 for
$c_0=0$. The curves have been obtained for the best fit values
indicated in Table \ref{tableFitBAOdz}. For more details, see
\cite{GoSolBas2014}. \label{GrowingRate}}
\end{center}
\end{figure}

%

It t is pretty clear from Fig.\,\ref{GrowingRate} that the dynamical
vacuum models under consideration are able to adjust the linear
growth data in a way comparable to the $\CC$CDM. This was expected
since the ``running'' of the vacuum density in our time does not
depart too much from the constant $\rLo$ value. However, if one sets
$c_0=0$ in the running models the departure from the structure
formation data becomes dramatic. This is quite manifest in the
anomalous behavior of the C1 curve in Fig.\,\ref{GrowingRate}.

We have performed the fit of the various vacuum models from the combined
data on type Ia supernovae (SNIa)\,\cite{Suzuki:2011hu}, the shift
parameter of the Cosmic Microwave Background
(CNB)\,\cite{CosmoObservations2}, and the data on the Baryonic
Acoustic Oscillations (BAOs)\,\cite{Blake11}. In the last case we
used the BAO$_{dz}$ data based on the $d_z$ estimator defined in
that reference -- confer  Table 3 in it. The basic fitting procedure
is explained in detail in Ref.\,\cite{GoSolBas2014}. Our results are
collected in Table 1 above. Owing to some degeneracies in the two
parameters of models A2 and B2, for the former we have set
$\tilde{\nu}=\nu/2$ (such that $\xiR=1$) and for the latter
$\nu=\epsilon$. With this setting, and taking into account that
$|\nu|,|\tilde{\nu}|\ll 1$,  we have from
(\ref{defxiM}) $\xi\simeq 1-\nueff$, with
$\nueff=\nu-\frac32\tilde{\nu}=\frac14\nu$. The fitting value to
this effective parameter is in fact the one quoted in Table 1 for
model A2.

Parameterizing the linear growth as $f(z)\simeq
\Omega_{m}(z)^{\gamma(z)}$, one can define the linear growth rate
index\,\cite{Peebles1993} $\gamma$. It can be used to distinguish
cosmological models, see e.g. \,\cite{Athina2014}. For the
$\Lambda$CDM model such index is approximated by $\gamma_{\CC}
\simeq 6/11\simeq 0.545$. For the dynamical vacuum models we make
use of (\ref{eq:growingfactor}) and the knowledge of the
corresponding growth factor $D(a)$ for each model, and we get
\begin{equation}\label{growthrateindex}
\gamma(z)\simeq \frac{\ln\left[-(1+z) \frac{d\ln D}{dz}\right]}
{\ln\Omega_{m}(z)} \;.
\end{equation}
The growth rate index $\gamma$ for each model is defined from the
value of the previous  expression for $z=0$.

\begin{figure}[!t]
\begin{center}
\includegraphics[scale=0.33]{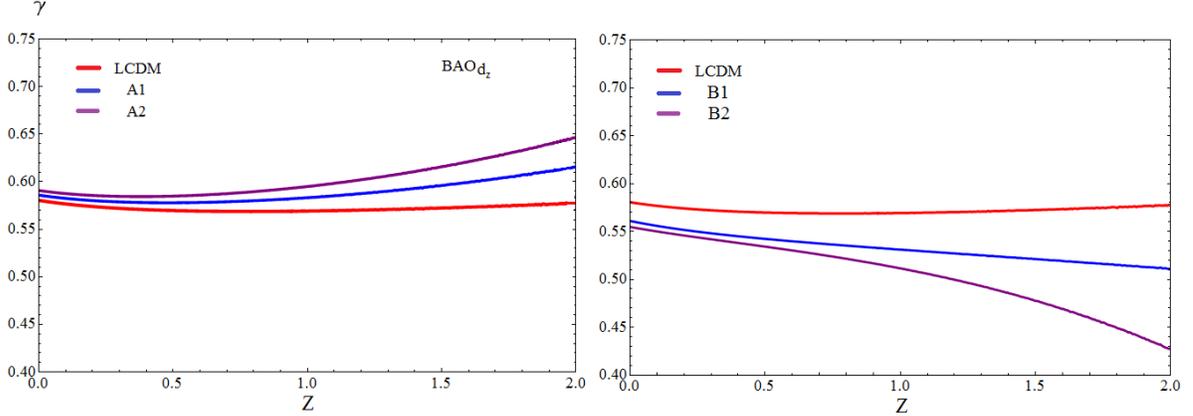}
\caption{\footnotesize{The evolution of the growth rate index,
Eq.\,(\ref{growthrateindex}), for the various vacuum models under the
same inputs as in the previous figure. \label{GrowthIndexTypeA}} }
\end{center}
\end{figure}

In Fig.\,\ref{GrowthIndexTypeA} we plot the evolution of $\gamma(z)$
for the A and B type of vacuum models. In the same figures we can
also see our determination of $\gamma_{\CC}(z)$ for the $\CC$CDM as
a function of the redshift, and in particular we find
$\gamma_{\CC}(0)\simeq 0.58$. From that figure we see that the
growth index of the dynamical vacuum models (especially for type-A models) is well approximated by
the $\Lambda$CDM constant value for $z<1$, while at large
redshifts there are deviations. It is worth mentioning that the
differences with respect to the $\CC$CDM are at the edge of the
present experimental limits. For example, in a recent analysis it is
found that $\gamma=0.56\pm 0.05$ and
$\Omo=0.29\pm0.01$\,\cite{Athina2014}. The prediction of $\gamma$
for all our vacuum models lies within $1\sigma$ of that range. Being
the differences on the verge of being observed, we expect that in
the future it should be possible to discriminate among the various $\bCDM$
models by this procedure.

%
%

Finally, we briefly mention an alternative procedure to distinguish the running vacuum models from the standard $\CC$CDM. It is based on the cluster number
counts method, see\,\cite{Grande2011,BPS09} for previous
applications in this context. The method is based on the
Press-Schechter formalism\,\cite{PS1974} and generalizations
thereof\,\cite{GeneralPS}. We refer the reader to
Refs.\,\cite{GoSolBas2014,SolGo2014} for a comprehensive presentation, and in what follows we show only the most important results. The main
observable of the method is the fractional difference $\delta {\cal
N}(z)/{\cal N}(z)$ in the number of counts of clusters between the
vacuum models and the concordance $\CC$CDM model at any given
redshift $z$. The number of counts at redshift $z$ is given by the
following expression:
\begin{equation}\label{eq:Nzb}
\mathcal{N}(z)=-\frac{4\pi
r^2(z)\,\bar{\rho}(z)}{H(z)}\,\int_{M_1}^{M_2}\frac{1}{M}\left(\frac{1}{\sigma}\frac{d\sigma}{dM}\right)f_{\rm PSc}(\sigma)dM\,.
\end{equation}
Here $\bar{\rho}(z)$ is the comoving background mass density. The original Press-Schechter function is\,\cite{PS1974} $f_{\rm PSc}(\sigma)=\sqrt{2/\pi} (\delta_c/\sigma)
\exp(-\delta_c^2/2\sigma^2)$. It depends on the parameter $\delta_{c}$, the linearly extrapolated density threshold above which structures collapse. This parameter must be computed for each vacuum model using the non-linear perturbations equations\,\cite{GoSolBas2014,SolGo2014}. A generalized and improved form of the Press-Schechter function
 $f_{\rm PSc}(\sigma)$, which we adopt here, is the
one provided by Reed et al. in \,\cite{GeneralPS}. Finally, $r(z)$ in the above equation is
the comoving radial distance out to redshift $z$, namely:
\begin{equation}\label{rzdef}
r(z)=\int_{0}^{z} \frac{dz'}{H(z')}\,;
\end{equation}

\noindent and $\sigma^2(M,z)$ is the mass variance
of the smoothed linear density field. It depends on the redshift
$z$ at which the halos are identified and is given by
\begin{equation}\label{sigma2}
\sigma^2(M,z)=\frac{D^2(z)}{2\pi^2} \int_0^\infty k^2 P(k) W^2(kR)
dk \,. \end{equation} In this expression, $P(k)$ is the power-spectrum and $D(z)$ is the linear
growth factor of perturbations. It is obtained from solving the
perturbations Eq.\,(\ref{diffeqD}) and is therefore characteristic
of each model.  Finally, $W$ is the Fourier transform of the standard
top-hat function with spherical symmetry -- for details, see e.g.
\,\cite{GoSolBas2014}.


\begin{figure}[!t]
\begin{center}
{\includegraphics[scale=0.55]{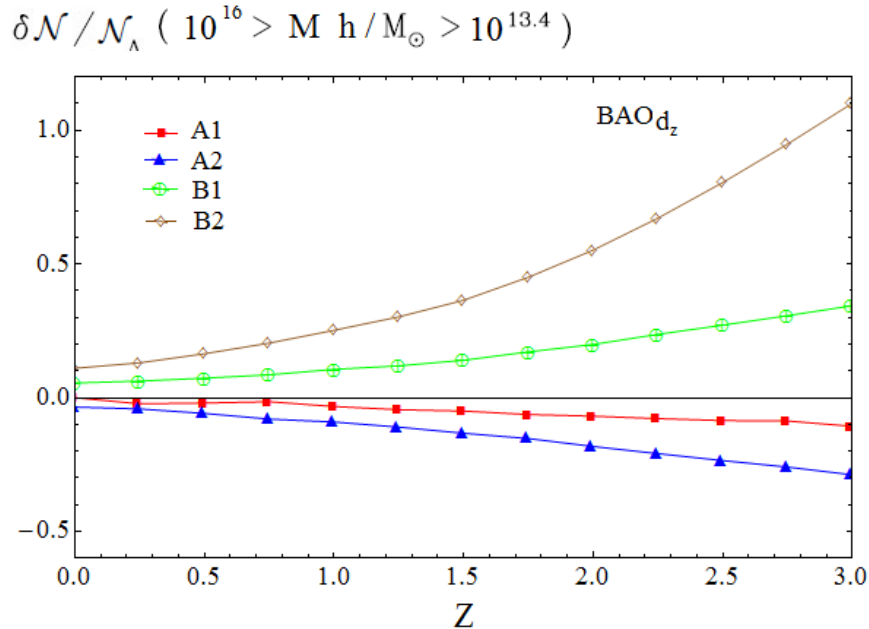}}
\end{center}
\caption{\scriptsize Comparison of the fractional difference
$\delta {\cal N}/{\cal N}$ in the redshift distribution of cluster
number counts (in the indicated range of masses) for the vacuum
models under consideration with respect to the concordance $\CC$CDM
model, i.e. $\delta {\cal N}={\cal N}-{\cal N}_{\CC CDM}$. Inputs as in the previous two figures.}
\label{NC:CentralFits}
\end{figure}


By selecting an appropriate redshift range for the observations, as
well as a characteristic range of masses $M_1<M<M_2$ for the
observed clusters, one can provide definite predictions from
(\ref{eq:Nzb}). Using the best fit values of the parameters of our
vacuum models, as  indicated in Table 1, and the full machinery of the
generalized Press-Schechter formalism, we can compute the redshift
distribution of clusters predicted in each vacuum model and subtract
the corresponding prediction from the $\CC$CDM, i.e. the quantity
$\delta {\cal N}(z)={\cal N}(z)-{\cal N}(z)_{\CC CDM}$. Then we can
compute the fractional difference $\delta {\cal N}(z)/{\cal
N}(z)_{\CC CDM}$, which encodes the relative deviations with respect to the concordance model. The numerical results are shown in
Fig.\,\ref{NC:CentralFits}.

The vacuum models are seen to be clearly separated by the cluster number
counts method, with the type-B ones providing an excess in the
number of counts, and the type-A models a defect, as compared to the
$\CC$CDM. The deviations can be significant ($+30-40\%$) for the former (especially for B2) but are moderate (say $-10\%$ to $-20\%$) for the
latter, at around the optimal redshif $z\simeq 1.5$ where it lies
(approximately) the maximum total number of counts (cf.
\cite{GoSolBas2014}). The fractional differences are still bigger
for higher redshifts, reaching $+50\%$ for B2 at $z\simeq 2$. Beyond
redshift $z=3$, however, the statistics of the total number of
counts depletes significantly and as a result this range is not
appropriate for practical measurements.

\section{Conclusions}

In this work we have discussed various aspects of the `` running
$\bCDM$ cosmology'' (a large class of dynamical vacuum models) in which the vacuum energy density can be expressed as a power series of the Hubble function and its cosmic
time derivatives. Some of these models can be well motivated within
the context of quantum field theory (QFT) in curved spacetime and
they can provide an overarching description of the cosmic history,
namely starting from the early inflationary times till our dark
energy (DE) days. For the study of the current Universe the series
naturally terminates at the level of the $H^2$ and $\dot{H}$ terms,
but the higher order ones  are crucially important for a proper description of
the inflationary phase.

The class of $\bCDM$ models involving higher powers of the Hubble
rate (typically $H^4$) have a non-singular starting point
characterized by an initial de Sitter period of rapid inflation,
presumably triggered by the high energy dynamics of a Grand Unified
Theory at a scale $M_X$ near (but below) the Planck scale. Explicit
solution of the model shows that the inflationary period is followed
by successful ``graceful exit'' into the standard radiation epoch.
The Universe continues into the cold dark matter era and finally
leads to a dark energy epoch with a very small amount of vacuum
energy.

It is important to emphasize that the post-inflationary period leads
to a cosmological regime that is very similar to the concordance
$\CC$CDM model, but with a distinctive feature: the vacuum energy is
not rigid but mildly dynamical ($\rL\sim c_0+\nu\,H^2$). This
feature can be tested and may prove a low-energy ``smoking gun'' of
the underlying vacuum dynamics. Let us remark the recent indications on dynamical DE in the current data\,\cite{DynamicalDE}. The slow vacuum evolution in our recent past, still going on in our days, could be responsible for the dynamical DE and could provide
an alternative explanation to it which is completely different\,\cite{GoSolBas2014} from
quintessence approaches and the like. Similarly, the same unified vacuum
structure explaining the DE is responsible, at the very early times,
of the inflationary period and without invoking at all any sort of
ad hoc inflaton fields. Therefore, in contrast to the $\CC$CDM
model, the very early times of the cosmic history in the $\bCDM$ cosmology is characterized by a powerful dynamical vacuum ($\sim H^4\sim M_X^4$),
which triggers inflation and upon decay into matter becomes much more moderate ($\sim H^2\gtrsim H_0^2$) near our present until effectively behaving as DE.

Quite noticeably, the running $\bCDM$ model provides a natural
explanation for the huge entropy of the current Universe. It also emerges from the primeval vacuum decay in the early Universe. While the details of the vacuum decay depend on the Grand Unified Theory  that brings about inflation, the final prediction of the entropy is universal and does not depend on the particular GUT implementation. The mechanism we have described here explains the value of the entropy in the current horizon.
No horizon problem exists at all in the $\bCDM$ model since all of the
points of the current Hubble sphere remain causally connected as of
the early times when a huge amount of relativistic particles emerged
out of the decaying dynamics of the primeval vacuum.

We have signaled some potential connection of the
$\sim H^4$ structure of the $\bCDM$ cosmology with Starobinsky
inflation, characterized by $R^2\sim H^4$.  The two sorts of models
present some similarities but they actually involve also important
differences, which depend on particular implementations.

After the inflationary epoch is left behind, the total amount of
vacuum energy decreases very fast, the primordial nucleosynthesis
can operate normally within a standard radiation epoch and the
Universe can go through the cold dark matter era until the present DE era. The latter appears here as a slowly time-varying vacuum-dominated epoch. The low-energy vacuum dynamics can, however, adopt different forms (which we have called type-A and type-B).

We have shown that at leading order all these forms  are
admissible structures for the consistent description of our
Universe, but they exhibit some differences that can be checked
observationally. In particular we have confronted our vacuum models
against the Hubble expansion data and structure formation. In
addition we have assessed their considerably different capability in populating
the Universe with virialized structures at different
redshifts as compared to the $\CC$CDM model.

The current Universe appears in all these models as FLRW-like,
except that the vacuum energy is not a rigid quantity but a mildly
evolving one. The typical values we have obtained for the
coefficients $\nu$, $\alpha$ and $\epsilon$ responsible for the time
evolution of $\rL$ lie in the ballpark of $\sim
10^{-3}$. This order of magnitude value is roughly consistent with
the theoretical expectations, some of them interpreted in QFT as
one-loop $\beta$-functions of the running cosmological
constant\,\cite{JSP-CCReview2013,Fossil07,SS0809}. It is  rather encouraging since it
points to a fundamental origin of the theoretical structure of the  $\bCDM$
models in the context of QFT in curved spacetime.

To summarize, the class of $\bCDM$ cosmologies may offer an appealing and phenomenologically consistent perspective for
describing inflation and dynamical dark energy without introducing
extraneous dark energy fields. Ultimately they might offer a
clue to better understand the origin of the $\CC$-term and the
cosmological constant problem in the context of fundamental physics.
It would be a timely achievement, if we take into consideration that we are
currently approaching the centenary of the introduction of the
cosmological term in Einstein's equations
in 1917\,\cite{Einstein1917}.

\newpage

{\bf Acknowledgments}
JS has been supported in part by FPA2013-46570 (MICINN), Consolider
grant CSD2007-00042 (CPAN) and by 2014-SGR-104 (Generalitat de
Catalunya). The work of AGV has been partially supported by an APIF
predoctoral grant of the Universitat de Barcelona. One of us (JS) is
thankful to S. Basilakos, H. Fritzsch, J.A.S. Lima, N.E. Mavromatos
and D. Polarski for the recent collaboration in some of the work
presented here.

\newcommand{\JHEP}[3]{ {JHEP} {#1} (#2)  {#3}}
\newcommand{\NPB}[3]{{ Nucl. Phys. } {\bf B#1} (#2)  {#3}}
\newcommand{\NPPS}[3]{{ Nucl. Phys. Proc. Supp. } {\bf #1} (#2)  {#3}}
\newcommand{\PRD}[3]{{ Phys. Rev. } {\bf D#1} (#2)   {#3}}
\newcommand{\PLB}[3]{{ Phys. Lett. } {\bf B#1} (#2)  {#3}}
\newcommand{\EPJ}[3]{{ Eur. Phys. J } {\bf C#1} (#2)  {#3}}
\newcommand{\PR}[3]{{ Phys. Rep. } {\bf #1} (#2)  {#3}}
\newcommand{\RMP}[3]{{ Rev. Mod. Phys. } {\bf #1} (#2)  {#3}}
\newcommand{\IJMP}[3]{{ Int. J. of Mod. Phys. } {\bf #1} (#2)  {#3}}
\newcommand{\PRL}[3]{{ Phys. Rev. Lett. } {\bf #1} (#2) {#3}}
\newcommand{\ZFP}[3]{{ Zeitsch. f. Physik } {\bf C#1} (#2)  {#3}}
\newcommand{\MPLA}[3]{{ Mod. Phys. Lett. } {\bf A#1} (#2) {#3}}
\newcommand{\CQG}[3]{{ Class. Quant. Grav. } {\bf #1} (#2) {#3}}
\newcommand{\JCAP}[3]{{ JCAP} {\bf#1} (#2)  {#3}}
\newcommand{\APJ}[3]{{ Astrophys. J. } {\bf #1} (#2)  {#3}}
\newcommand{\AMJ}[3]{{ Astronom. J. } {\bf #1} (#2)  {#3}}
\newcommand{\APP}[3]{{ Astropart. Phys. } {\bf #1} (#2)  {#3}}
\newcommand{\AAP}[3]{{ Astron. Astrophys. } {\bf #1} (#2)  {#3}}
\newcommand{\MNRAS}[3]{{ Mon. Not. Roy. Astron. Soc.} {\bf #1} (#2)  {#3}}
\newcommand{\JPA}[3]{{ J. Phys. A: Math. Theor.} {\bf #1} (#2)  {#3}}
\newcommand{\ProgS}[3]{{ Prog. Theor. Phys. Supp.} {\bf #1} (#2)  {#3}}
\newcommand{\APJS}[3]{{ Astrophys. J. Supl.} {\bf #1} (#2)  {#3}}

\newcommand{\Prog}[3]{{ Prog. Theor. Phys.} {\bf #1}  (#2) {#3}}
\newcommand{\IJMPA}[3]{{ Int. J. of Mod. Phys. A} {\bf #1}  {(#2)} {#3}}
\newcommand{\IJMPD}[3]{{ Int. J. of Mod. Phys. D} {\bf #1}  {(#2)} {#3}}
\newcommand{\GRG}[3]{{ Gen. Rel. Grav.} {\bf #1}  {(#2)} {#3}}



\end{document}